\documentclass[runningheads,fleqn]{svmult}

\usepackage{makeidx}   
\usepackage{epsfig}  
\usepackage{subeqnar}  
\usepackage{multicol}  
\usepackage{physmult}  
\makeindex             

\def\lsim{\raise0.3ex\hbox{$<$\kern-0.75em\raise-1.1ex\hbox{$\sim$}}} 
\def\gsim{\raise0.3ex\hbox{$>$\kern-0.75em\raise-1.1ex\hbox{$\sim$}}}

\newcommand{\plaq}{\mbox{\raisebox{-1.25mm}
{\epsfig{file=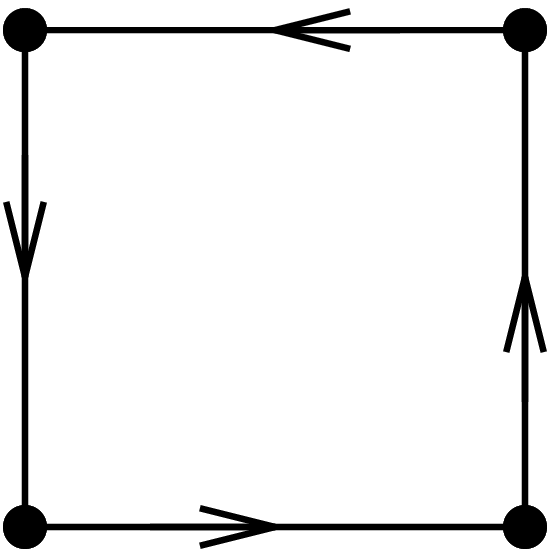,height=5mm
}}~}}
\newcommand{\loOp}{\mbox{\raisebox{-1.25mm}
{\epsfig{file=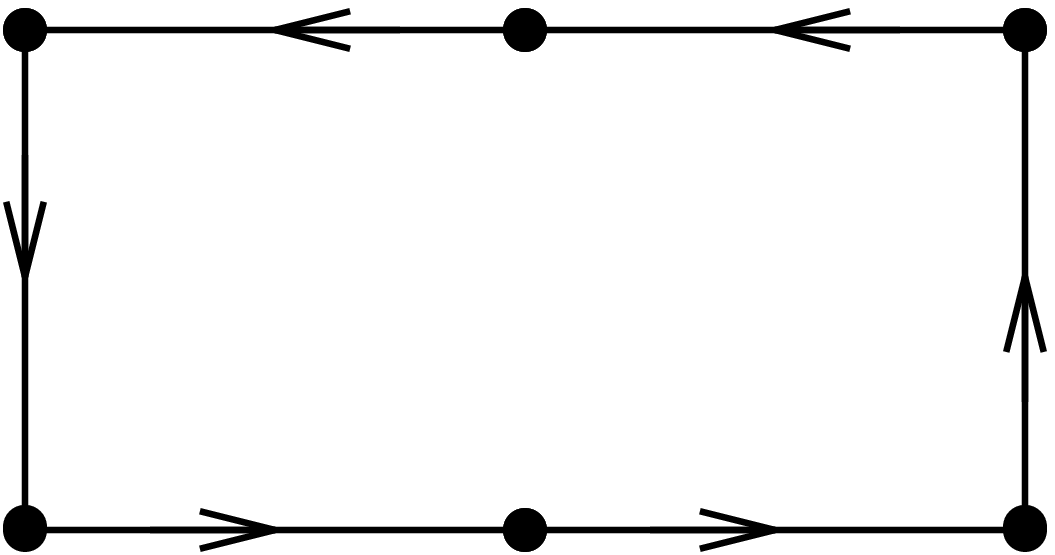,height=5mm
}}~}}
\newcommand{\lOop}{\mbox{\raisebox{-4mm}
{\epsfig{file=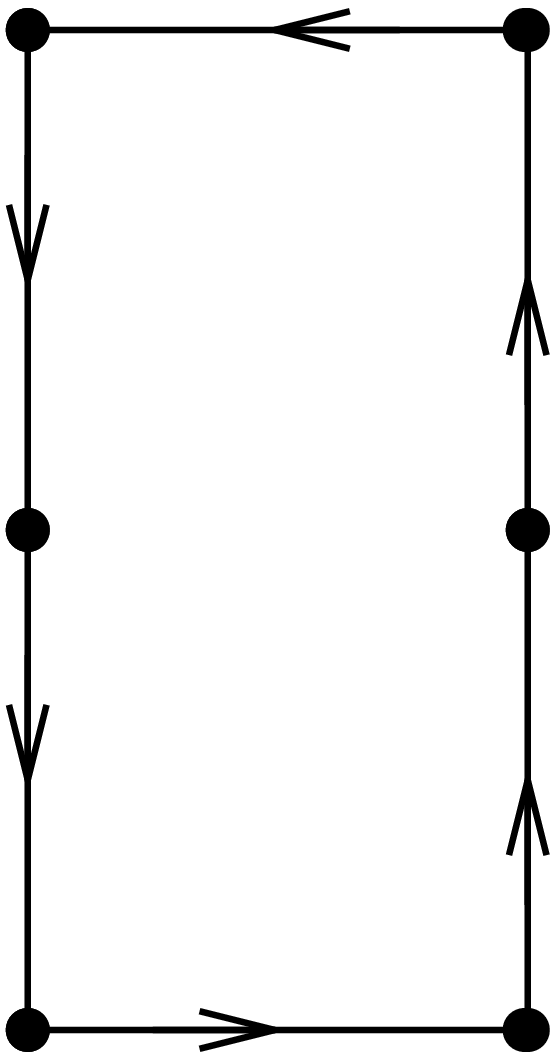,height=10mm
}}~}}
\newcommand{\alink}{\mbox{
\begin{picture}(2.5,.2)
\linethickness{1mm}
\multiput(0,0.1)(2,0){2}{\circle*{0.1}}
\multiput(1.01,0.1)(0,0){1}{\circle{0.2}}
\put(0.8,0.){\link}
\put(-0.2,0.){\linka}
\put(1.1,-.2){\scriptsize \( x \) }
\put(0,-.2){\scriptsize \( y \) }
\put(2.0,-.2){\scriptsize \( y \) }
\end{picture}}}
\newcommand{\alinkfat}{\mbox{
\begin{picture}(3.0,.2)
\linethickness{1mm}
\multiput(0,0.1)(2.4,0){2}{\circle*{0.1}}
\multiput(1.21,0.1)(0,0){1}{\circle{0.2}}
\put(1.2,-1.){\staple}
\put(-0.2,-1.){\staplea}
\put(1.1,-.3){\scriptsize \( x \) }
\put(-.4,-.2){\scriptsize \( y \) }
\put(2.6,-.2){\scriptsize \( y \) }
\end{picture}}}
\newcommand{\blinkba}{\mbox{
\begin{picture}(2.5,2.5)
\thicklines
\multiput(1,-1)(0,-1){2}{\circle*{0.1}}
\multiput(1,0)(0,0){1}{\circle{0.2}}
\multiput(2,0.0)(0,1){3}{\circle*{0.1}}
\multiput(0,-2)(0,0){1}{\circle*{0.1}}
\put(1,-2.0){\vector(-1,0){1}}
\put(1,-1.0){\vector(0,-1){1}}
\put(1,0.0){\vector(0,-1){1}}
\put(1,0.0){\vector(1,0){1}}
\put(2,0.0){\vector(0,1){1}}
\put(2,1.0){\vector(0,1){1}}
\put(0.5,-0.1){\scriptsize \( x \) }
\put(1.5,2.0){\scriptsize \( y \) }
\put(-0.4,-2.1){\scriptsize \( y \) }
\end{picture}}}
\newcommand{\blinkbb}{\mbox{
\begin{picture}(2.5,2.5)
\thicklines
\multiput(0,0.0)(0,-1){3}{\circle*{0.1}}
\multiput(1,1.0)(0,1){2}{\circle*{0.1}}
\multiput(1,0.0)(0,){1}{\circle{0.2}}
\multiput(2,2)(0,1){1}{\circle*{0.1}}
\put(0,-1.0){\vector(0,-1){1}}
\put(0,0){\vector(0,-1){1}}
\put(1,0){\vector(-1,0){1}}
\put(1,0){\vector(0,1){1}}
\put(1,1){\vector(0,1){1}}
\put(1,2){\vector(1,0){1}}
\put(1.25,-0.1){\scriptsize \( x \) }
\put(2.2,2.0){\scriptsize \( y \) }
\put(-0.4,-2.1){\scriptsize \( y \) }
\end{picture}}}
\newcommand{\blinkbc}{\mbox{
\begin{picture}(2.5,2.5)
\thicklines
\multiput(1,1)(0,1){2}{\circle*{0.1}}
\multiput(1,0)(0,0){1}{\circle{0.2}}
\multiput(2,0)(0,-1){3}{\circle*{0.1}}
\multiput(0,2)(0,1){1}{\circle*{0.1}}
\put(1,2){\vector(-1,0){1}}
\put(1,1){\vector(0,1){1}}
\put(1,0){\vector(0,1){1}}
\put(1,0){\vector(1,0){1}}
\put(2,0){\vector(0,-1){1}}
\put(2,-1){\vector(0,-1){1}}
\put(0.5,-0.1){\scriptsize \( x \) }
\put(1.5,-2.1){\scriptsize \( y \) }
\put(-0.5,2.0){\scriptsize \( y \) }
\end{picture}}}
\newcommand{\blinkbd}{\mbox{
\begin{picture}(2.5,2.5)
\thicklines
\multiput(0,0)(0,1){3}{\circle*{0.1}}
\multiput(1,-1)(0,-1){2}{\circle*{0.1}}
\multiput(1,0)(0,-1){1}{\circle{0.2}}
\multiput(2,-2)(0,1){1}{\circle*{0.1}}
\put(0,1){\vector(0,1){1}}
\put(0,0){\vector(0,1){1}}
\put(1,0){\vector(-1,0){1}}
\put(1,0){\vector(0,-1){1}}
\put(1,-1){\vector(0,-1){1}}
\put(1,-2){\vector(1,0){1}}
\put(1.25,-0.1){\scriptsize \( x \) }
\put(2.1,-2.1){\scriptsize \( y \) }
\put(-0.5,2.0){\scriptsize \( y \) }
\end{picture}}}
\newcommand{\staple}{\mbox{
\begin{picture}(1.2, 2.2)
\thicklines
\put(0,1.2){\vector(0,1){1}}
\put(0,1){\vector(0,-1){1}}
\put(0,2.2){\vector(1,0){1}}
\put(0,0.){\vector(1,0){1}}
\put(1,0.){\vector(0,1){1}}
\put(1,2.2){\vector(0,-1){1}}
\end{picture}}}
\newcommand{\staplea}{\mbox{
\begin{picture}(1.2, 2.2)
\thicklines
\put(1,1.2){\vector(0,1){1}}
\put(0,2.2){\vector(0,-1){1}}
\put(1,2.2){\vector(-1,0){1}}
\put(1,1){\vector(0,-1){1}}
\put(1,0){\vector(-1,0){1}}
\put(0,0){\vector(0,1){1}}
\end{picture}}}
\newcommand{\link}{\mbox{
\begin{picture}(1.1, .1)
\thicklines
\put(0,0.1){\vector(1,0){1}}
\end{picture}}}
\newcommand{\linka}{\mbox{
\begin{picture}(1.1, .1)
\thicklines
\put(1,0.1){\vector(-1,0){1}}
\end{picture}}}

\begin{document}
\title*{Lattice QCD at High Temperature and Density}
\toctitle{Lattice QCD at High Temperature and Density}
%
%
\titlerunning{Lattice QCD at High Temperature and Density}
%
\author{Frithjof Karsch, \\
Fakult\"at f\"ur Physik, Universit\"at
Bielefeld, D-33615 Bielefeld, Germany}
\authorrunning{Frithjof Karsch}
%
%
\maketitle              

\footnotesize
\noindent
{\bf Abstract:} After a brief introduction into basic aspects of
the formulation of lattice regularized QCD at finite temperature
and density we discuss our current understanding of the QCD
phase diagram at finite temperature. We present results from
lattice calculations that emphasize the deconfining as well
as chiral symmetry restoring features of the  QCD transition,
and discuss the thermodynamics of the high temperature phase.

\section{Introduction}
 
\vskip -250pt
{\small
\mbox{} \hfill BI-TP 2001/10\\
\mbox{} \hfill June 2001\\
}
\vskip 225pt

Almost immediately after the ground-breaking demonstration that 
the numerical analysis of lattice regularized quantum field theories
\cite{Wilson}
can also provide quantitative results on fundamental non-perturbative
properties of QCD \cite{Creutz} it has been realized that this approach
will also allow to study the QCD phase transition
\cite{McLerran,Kuti} and the equation of state of the quark-gluon
plasma \cite{Bielefeld}. During the last 20 years we have learned a 
lot from lattice calculations about the phase structure of QCD at finite 
temperature. In fact, we do understand quite well the thermodynamics
in the heavy quark mass limit of QCD, the pure $SU(3)$ gauge theory,
and even have calculated the critical temperature and the equation
of state in this limit with an accuracy of a few percent. However, it 
is only now that we start to reach a level of accuracy in numerical 
calculations of QCD thermodynamics that allows to seriously consider 
quantitative studies of QCD with a realistic light quark mass spectrum. 
An important ingredient in the preparation of such calculations
is the development of new regularization schemes in the fermion
sector of the QCD Lagrangian, which allow to reduce discretization
errors and also improves the flavour symmetry of the lattice actions.
The currently performed investigations of QCD thermodynamics  
provide first results with such improved actions and prepare 
the ground for calculations with a realistic light quark mass 
spectrum. 

The interest
in analyzing the properties of QCD under extreme conditions is twofold.
On the one hand it is the goal to reach a quantitative description
of the behaviour of matter at high temperature and density. This does
provide important input for a quantitative description of experimental
signatures for the occurrence of a phase transition in heavy ion
collisions and should also help to understand better the phase
transitions that
occurred during the early times of the evolution of the universe.
Eventually it also may allow to answer the question whether a
quark-gluon
plasma can exist in the interior of dense neutron stars or did exist in
early stages of supernova explosions.
For this reason one would like
to reach a quantitative understanding of the QCD equation of state,
determine critical parameters such as the critical temperature and
the critical energy density and predict the modification of basic
hadron properties (masses, decay widths) with temperature.
On the other hand the analysis of a complicated quantum field theory
like QCD at non-zero temperature can also help to improve our understanding
of its non-perturbative properties at zero temperature. The introduction
of external control parameters (temperature, chemical potential) allows 
to observe the response of different observables to this and may provide 
a better understanding of their interdependence \cite{Wilczek}.
As such one would, for instance, like to clarify
the role of confinement and chiral symmetry
breaking for the QCD phase transition. In which respect is the QCD
phase transition deconfining and/or chiral symmetry restoring? In
how far can the critical behaviour be described by intuitive pictures
based on percolation, bag or resonance gas models which have been 
developed for the QCD transition? We will discuss these qualitative
aspects of the QCD thermodynamics and also present results on
basic questions concerning the equation of state and the critical
temperature of the transition which ask for quantitative answers.

In the next section we give a short introduction into the 
lattice formulation of QCD thermodynamics. In Section 3 we discuss
the basic structure of the QCD phase diagram at finite temperature
as it is known from lattice calculations. Section 4 is devoted to
a discussion of basic thermodynamic observables which characterize
the QCD transition to the plasma phase and we will identify general 
properties which show the deconfining and chiral symmetry restoring 
features of this transition. In Section 5 we comment on different
length scales characterizing the QCD plasma and try to establish the
temperature regime where lattice calculations may make contact with
perturbative approaches. A description of recent results on the QCD 
equation of state and the critical temperature of the QCD transition
which emphasizes the quark mass and flavour dependence of these
quantities is given in Sections 6 and 7, respectively. A brief discussion
of the problems arising in lattice formulations of QCD at non-zero
baryon number density or chemical potential is given in Section 8.
Finally we give our conclusions in Section 9 and describe a specific set 
of improved gauge and fermion lattice actions in an Appendix.

\section{The Lattice Formulation of QCD Thermodynamics}

\subsection{The basic steps from continuum to lattice ...}

Starting point for the discussion of the equilibrium thermodynamics of
QCD on the lattice is the QCD partition function, which explicitly
depends on the volume ($V$), the temperature ($T$) and the quark number
chemical potential ($\mu$). It is represented in terms of a Euclidean
path integral over gauge ($A_\nu$) and fermion ($\bar{\psi},\; \psi$)
fields,

\begin{equation}
Z (V,T,\mu) =
\int \;{\cal D} A_\nu {\cal D}\bar{\psi}{\cal D}\psi\;
{\rm e}^{-S_E(V,T,\mu)} \quad . 
\label{partZ}
\end{equation}
where $A_\nu$ and $\bar{\psi},\; \psi$ obey periodic and anti-periodic 
boundary conditions in Euclidean time, respectively. The Euclidean
action $S_E\equiv S_G +S_F$ contains a purely gluonic contribution
($S_G$) expressed in terms of the field strength tensor, 
$F_{\mu\nu} = \partial_\mu A_\nu - \partial_\nu A_\mu - \I g
[A_\mu,A_\nu]$, and a fermion part ($S_F$), which couples the gauge and
fermions field through the standard minimal substitution, 
\begin{eqnarray}
S_E(V,T,\mu) &\equiv& S_G(V,T)\; + \; S_F(V,T,\mu) \\
S_G(V,T)&=& \int\limits_0^{1/T} \D x_0 \int\limits_V \D^3 {\bf x} \;
\frac{1}{2} {\rm Tr}\; F_{\mu\nu} F_{\mu\nu}  \\
S_F(V,T,\mu)&=& \int\limits_0^{1/T} \D x_0 \int\limits_V \D^3 {\bf x} \;
 \sum_{f=1}^{n_f} \bar{\psi}_f \left( \gamma_\mu
    [\partial_\mu-\I g A_\mu] + m_f- \mu \gamma_0 \right) \psi_f .
\label{lagrangian}
\end{eqnarray}
Here $m_f$ are the different quark masses for the $n_f$ different 
quark flavours and $g$ denotes the QCD coupling constant.

The path integral appearing in Eq.~\ref{partZ} is regularized by introducing 
a four dimensional space-time lattice of size $N_\sigma^3 \times N_\tau$
with a lattice spacing $a$. Volume and temperature are then related
to the number of points in space and time directions, respectively,
\begin{equation}
  V= (N_\sigma\; a)^3 \quad, \quad T^{-1} = N_\tau\; a\quad .
\end{equation} 
While the discretization of the fermion sector, at least on the naive
level, is straightforwardly achieved by replacing derivatives by
finite differences, the gauge sector is a bit more involved. Here
we introduce {\it link variables} $U_\mu (x)$ which are associated with
the link between two neighbouring sites of the lattice and describe
the parallel transport of the field $\cal A$ from site $x$ to
$x+\hat{\mu}a$,
\begin{equation}
\label{link}
U_{x,\mu} = {\rm P} \exp\biggl( i g \int_x^{x+\hat{\mu} a} \D x^\mu\, 
A_\mu(x)\biggr) \quad ,
\end{equation}
where $\rm P$ denotes the path ordering. The link variables $U_\mu (x)$
are elements of the $SU(3)$ colour group. A product of these
link variables around an elementary plaquette may be used to define
an approximation to the gauge action,
\begin{eqnarray}
W_{n,\mu\nu}^{(1,1)} = 1 - \frac{1}{3}{\rm Re}\; \plaq_{n,\mu \nu} &\equiv& 
{\rm Re\; Tr}\; U_{n,\mu}U_{n+\hat{\mu},\nu}
U_{n+\hat{\nu},\mu}^{\dagger} U_{n,\nu}^{\dagger} \nonumber \\
&=& \frac{g^2 a^4}{2} F_{\mu\nu}^a F_{\mu\nu}^a + {\cal O}(a^6)
\quad .
\end{eqnarray}
A discretized version of the Euclidean gauge action, which reproduces
the continuum version up to cut-off errors of order $a^2$, thus is
given by the {\it Wilson action} \cite{Wilson},
\begin{equation}
\beta S_G = \beta \sum_{\scriptstyle n \atop\scriptstyle 0\le\mu<\nu\le3}
W_{n,\mu\nu}^{(1,1)} \quad  \Longrightarrow\quad  \int\D^4 x\; 
{\cal L}_E \; +\; {\cal O}(a^2)
\quad ,
\end{equation}
where we have introduced the gauge coupling $\beta = 6 /g^2$.

As is well-known the naive discretization of the fermionic part of
the action, which is obtained by introducing the simple finite
difference scheme to discretize the derivative appearing in the
fermion Lagrangian, {\it i.e.} 
$\partial_\mu \psi_f(x) = (\psi_{n+\hat{\mu}}-\psi_{n-\hat{\mu}})/2a$, 
does in the continuum limit not reproduce the particle content one 
started with. The massless lattice fermion propagator has poles not 
only at zero momentum but also at all other corners of the 
Brillouin zone and thus generates 16 rather than a single fermion
species in the continuum limit. One thus faces a severe species doubling 
problem. The way out has been to either introduce an explicit chiral 
symmetry breaking term, which is proportional to $a\partial_\mu^2 \psi_f(x)$
and thus vanishes in the continuum limit (Wilson fermions \cite{Wilson}), 
or to distribute the components of the fermion Dirac spinors over several 
lattice sites (staggered fermions) \cite{KoSu}.
The staggered fermion formulation does not eliminate the species doubling 
problem completely. One still gets four degenerate fermion species.
However, it has the advantage that it preserves a continuous
subgroup of the original global chiral symmetry. In the massless limit
the chiral condensate thus still is an order parameter for the occurrence 
of a phase transition at finite temperature.

Progress has been made in formulating lattice QCD also with chiral 
fermion actions which do avoid the species doubling 
and at the same time preserve the chiral symmetry of the QCD Lagrangian. 
This can, for instance, be achieved by introducing an extra
fifth dimension \cite{Kaplan}. At
present, however, very little has been done to study QCD thermodynamics
on the lattice with these actions \cite{Columbia}. Much more is known on 
the QCD thermodynamics from calculations with Wilson and staggered fermions.
We will in the following present results from both approaches.
However, to be specific we will restrict ourselves here to a discussion
of the staggered fermion formulation introduced by Kogut and 
Susskind \cite{KoSu}. The fermion action can be written as
\begin{equation}
\label{staggered}
S_{F}^{KS}
= \sum_{n m} \bar{\chi}_n Q^{KS}_{n m} {\chi}_m \quad ,
\end{equation}  
where the staggered fermion matrix $Q^{KS}$ is given by
\begin{eqnarray}
\label{QF}
Q^{KS}_{n m}(m_q, \tilde{\mu}) &=& 
{1\over 2} \sum_{\mu =1}^3 (-1)^{n_0+\dots+n_{\mu-1}}
( \delta_{n+\hat{\mu},m} U_{n,\mu} - \delta_{n,m+\hat{\mu}}
  U_{m,\mu}^{\dagger}) \nonumber \\
& &+{1\over 2}(\delta_{n+\hat{0},m} U_{n,0}\; \E^{\tilde{\mu}} - 
\delta_{n,m+\hat{0}} U_{m,0}^{\dagger})\; \E^{-\tilde{\mu}} \; 
+ \; \delta_{n m} m_q \quad .
\end{eqnarray}
Here we have introduced the chemical potential $\tilde{\mu}$ on the
temporal links \cite{Hasenfratz}. As the fermion action
is quadratic in the Grassmann valued quark fields $\bar{\chi}$ and
$\chi$ we can integrate them out in the partition function and
finally arrive at a representation of $Z(V,T,\mu)$ on a 4-dimensional
lattice of size $N_\sigma^3 \times N_\tau$,

\begin{equation}
\label{zsf}
Z(N_\sigma, N_\tau, \beta, m_q, \tilde{\mu}) = 
\int \prod_{n \nu} \D U_{n,\nu} (\det Q^{KS}(m_q, \tilde{\mu}))^{n_f/4}
\E^{-\beta S_{G}} \quad .
\end{equation}
We have made explicit the fact that the staggered fermion action
does lead to four degenerate fermion flavours in the continuum
limit, {\it i.e.} taking the continuum limit with the action
given in Eqs.~\ref{staggered} and \ref{QF} corresponds to $n_f=4$
in Eq.~\ref{zsf}. As the number of fermion species does appear 
only as an appropriate power of the fermion determinant, which is true
also in the continuum limit, one also may choose $n_f \ne 4$ in 
Eq.~\ref{zsf}. This is the approach used to perform simulations 
with different number of flavours in the staggered fermion formulation.

For $\tilde{\mu} = 0$ the fermion determinant appearing in Eq.~\ref{zsf}
is real and positive. Standard numerical techniques, which rely on a
probability interpretation of the integrand in Eq.~\ref{zsf}, thus
can be applied. For $\tilde{\mu} \ne 0$ the determinant, however,
becomes complex. Although the contribution of the imaginary part can
easily be shown to be zero, as it should to give a real partition
function, the real part is no longer strictly positive. This {\it sign
problem} so far still constitutes a major problem in the application of
numerical techniques to studies of QCD at non-zero baryon number
density or non-zero chemical potential. We therefore will restrict our
discussion of QCD thermodynamics mainly to the case $\tilde{\mu} \equiv
0$ and will come back to the problems one faces for $\tilde{\mu} \ne 0$
in Section 8.

\subsection{... and back from the lattice to the continuum}

The lattice discretized QCD action discussed above reproduces the
continuum action up to discretization errors of ${\cal O} (a^2)$.
In order to perform the continuum limit at constant temperature,
we will have to take the limit 
$(a\rightarrow 0,\; N_\tau \rightarrow \infty)$ with $T=1/N_\tau a$ 
fixed. In particular, for bulk thermodynamic observables like
the pressure and energy density, which have dimension $[T^4]$
this limit is rather cumbersome. All lattice observables are
dimensionless and are thus calculated in appropriate units of
the lattice spacing $a$. As a consequence a calculation of, e.g., 
the pressure will provide $pa^4$ and thus yields a numerical result
which decreases in magnitude like $N_\tau^{-4}$. Numerical 
calculation, however, are always based on the analysis of a finite
set of suitably generated gauge field configurations and thus
produce results which have a statistical error. It therefore 
rapidly becomes difficult to calculate bulk thermodynamic 
quantities on lattices with large temporal extent $N_\tau$.
For this reason it is of particular importance for finite
temperature calculations to be able to use actions which
have small discretization errors and thus allow to perform 
calculations on lattices with moderate temporal extent. Such 
actions have been developed and successfully applied in thermodynamic
calculations for the pure $SU(3)$ gauge theory. In the
fermion sector appropriate actions, which reduce cut-off 
effects in the high temperature ideal gas limit, so far have 
only been constructed for staggered fermions. As an example we 
describe a specific choice of improved gauge and staggered fermion  
actions in more detail in an Appendix. 

As mentioned above we have to perform the continuum limit in order
to eliminate lattice discretization errors and to arrive finally
at quantitative predictions for the QCD thermodynamics.
Eventually we thus have to analyze our observables on different
size lattices and extrapolate our results 
to $N_\tau \rightarrow \infty$ at fixed temperature. 
Unless we perform calculations at a well defined temperature,
e.g. the critical temperature, we will have to determine the 
temperature scale from an additional (zero-temperature)
calculation of an observable for which we know its physical 
value (in MeV). This requires a calculation at the same value of the
cut-off (same values of the bare couplings).
Of course, we know such a quantity only for the physical case 
realized in nature, {\it i.e.} QCD with two light up and down quark
flavours and a heavier strange quark. Nonetheless, we have good 
reason to believe that certain observables are quite insensitive
to changes in the quark masses, e.g. quenched hadron masses\footnote{A
physical observable $O$ is calculated on the lattice as dimensionless quantity,
which we denote here by $\tilde{O}$. Quite often, however, we will also
adopt the customary lattice notation, which explicitly specifies
the cut-off dependence in the continuum limit, e.g. $\tilde{m}_H \equiv 
m_Ha$ or $\tilde{\sigma}\equiv \sigma a^2$.} ($\tilde{m}_H$)
or the string tension ($\tilde{\sigma}$) are believed to be suitable 
observables to set a physical scale even in the limit of infinite 
quark masses (pure $SU(3)$ gauge theory). We thus may use calculations
of these quantities to define a temperature scale,
\begin{equation}
T/\sqrt{\sigma} = 1/\sqrt{\tilde{\sigma}}N_\tau \quad {\rm or} 
\quad T/m_H = 1/\tilde{m}_H N_\tau \quad .
\label{tscale}
\end{equation} 
In the pure $SU(3)$ gauge theory as well as in the massless limit
the lattice spacing is controlled through $\beta$, the only bare 
coupling appearing in the Euclidean action. Asymptotically
$a$ and $\beta$ are then related through the leading order 
renormalization group equation,
\begin{equation}
\label{RGeq}   
a \Lambda_{L} \simeq (6b_0/\beta)^{-b_1/2 b_0^2} 
\E^{-\beta/12 b_0} \quad , 
\end{equation} 
where the two universal coefficients are given by,
\begin{equation}
\label{b0b1}
b_0 = \frac{1}{16\pi^2} \left( 11 - \frac{2}{3}
  n_f\right) \quad, \quad 
b_1 = \left(\frac{1}{16\pi^2}\right)^2 \left[ 102 
  - \left( 10 + \frac{8}{3} \right) n_f \right] \quad ,
\end{equation} 
and $\Lambda_L$ is a scale parameter which unambiguously can be
related to the scale parameter in other regularization schemes,
e.g. to $\Lambda_{\overline{MS}}$. The continuum limit thus is 
reached with increasing $\beta$.

In the case of non-zero quark masses one has in addition to insure 
that the continuum limit is taken along a {\it line of constant 
physics}. This can be achieved by keeping a ratio of hadron 
masses, for instance 
the ratio of pseudo-scalar and vector meson masses, $m_{PS}/m_V$,
constant while varying the couplings $(\beta, m_q)$. In the limit
$\beta \rightarrow \infty$ this requires a tuning of the bare
quark masses such that $m_q \rightarrow 0$. We also note that 
for small quark masses the vector meson mass $m_V$ approaches 
a constant, $m_V = m_\rho + {\cal O}(m_q)$, while the pseudo-scalar
is the Goldstone-particle corresponding to the broken chiral
symmetry of QCD (pion). Its mass is proportional to the square 
root of $m_q$.  
In the following we will quite often quote results as a function
of $m_{PS}/m_V$ which is just another way for quoting
results for different values of the quark mass. 

\section{The QCD Phase Diagram at Finite Temperature}
 
At vanishing baryon number density (or zero chemical potential)
the properties of the QCD phase transition depend on the number 
of quark flavours 
and their masses. While it is a detailed quantitative question
at which temperature the transition to the high temperature
plasma phase occurs, we do expect that the nature of the transition,
e.g. its order and details of the critical behaviour, are controlled
by global symmetries of the QCD Lagrangian. Such symmetries only
exist in the limits of either infinite or vanishing quark masses. For
any non-zero, finite value of quark masses the global symmetries
are explicitly broken. In fact, in the case of QCD the explicit 
symmetry breaking induced by the finite quark masses is very much 
similar to that induced by an external ferromagnetic field in spin 
models. We thus expect that a continuous phase transition, which may 
exist in the zero or infinite quark mass limit, will turn into a
non-singular crossover behaviour for any finite value of the 
quark mass. First order transitions, on the other hand, may persist
for some time before they end in a continuous transition.
Whether a true phase transition exists in QCD with the physically
realized spectrum of quark masses or whether in this case the 
transition is just a (rapid) crossover, again becomes a quantitative 
question which we have to answer through direct numerical calculations.

Our current understanding of the qualitative aspects of QCD phase 
diagram is based on universality arguments for the symmetry
breaking patterns in the heavy \cite{Yaffe} as well as the light 
quark mass regime \cite{Pis84,Gav94}. 
In the limit of infinitely heavy quarks, the pure $SU(3)$ gauge theory,
the large distance behaviour of the heavy quark free energy, $F_{\bar{q}q}$, 
provides a unique distinction between confinement
below $T_c$ and deconfinement for  $T \; >\; T_c$.
On a lattice of size $N_\sigma^3 \times N_\tau$ the heavy quark free
energy\footnote{In the $T\rightarrow 0$ limit this is just 
the heavy quark potential; at non-zero temperature $F_{\bar{q}q}$
does, however, also include a contribution resulting from the overall change
of entropy that arises from the presence of external quark and
anti-quark sources.} 
can be calculated from the expectation value of the Polyakov loop 
correlation function
\begin{equation}
\exp{\biggl( -{F_{\bar{q}q} (r,T) \over T}\biggr) } = \langle
{\rm Tr} L_{\vec{x}} {\rm Tr} L^{\dagger}_{\vec{y}} \rangle \quad,\quad
rT=|\vec{x}-\vec{y}| N_\tau
\label{poly}
\end{equation}
where $ L_{\vec{x}}$ and $L^{\dagger}_{\vec{y}}$  represent static
quark and anti-quark sources located at the spatial
points $\vec{x}$ and $\vec{y}$, respectively,
\begin{equation} 
L_{\vec{x}} = \prod_{x_0=1}^{N_\tau} U_{n,0} \quad , \quad
n\; \equiv\; (x_0, \vec{x}) \quad .
\label{polyakovloop}
\end{equation}
For large separations ($r\rightarrow \infty$) the correlation function
approaches $|\langle L \rangle|^2$, where 
$\langle L \rangle = N_{\sigma}^{-3} \langle \sum_{\vec{x}} {\rm
Tr}L_{\vec{x}} \rangle$ denotes the Polyakov loop expectation value, 
which therefore characterizes the behaviour of the heavy quark free energy 
at large distances and is an order parameter for deconfinement in
the $SU(3)$ gauge theory, 
\begin{equation}
\langle L \rangle \; \cases{=\; 0 \; \Leftrightarrow \;{\rm confined~ phase,}&
$T\; < \; T_c$ \cr
>\; 0 \; \Leftrightarrow \;{\rm deconfined~ phase,}& $T\; >\; T_c$}\quad .
\label{lorder}
\end{equation}
The effective theory for the order parameter is a 3-dimensional
spin model with global $Z(3)$ symmetry. Universality arguments then 
suggest that the phase transition is first order in the infinite
quark mass limit \cite{Yaffe}.

In the limit of vanishing quark masses the classical QCD Lagrangian is
invariant under chiral symmetry transformations; for $n_f$ massless
quark flavours the symmetry is
\vspace{0.2cm}

\centerline{$\displaystyle{U_A(1) \times SU_L(n_f) \times SU_R(n_f)}$.}

\vspace{0.2cm}
\noindent
However, only the $SU(n_f)$ flavour part of this symmetry is spontaneously
broken in the vacuum, which gives rise to $(n_f^2-1)$ massless Goldstone
particles, the pions. 
The axial $U_A(1)$ only is a symmetry of the classical Lagrangian.
It is explicitly broken due to quantum corrections in the QCD partition 
function, the axial anomaly, and therefore gets replaced by a discrete 
$Z(n_f)$ symmetry at low temperature.
The basic observable which reflects the chiral properties of QCD is 
the chiral condensate, 
\begin{equation}
\langle \bar{\chi}\chi  \rangle = {1\over N_\sigma^3 N_\tau}
{\partial \over \partial m_q}
\ln Z \quad .
\end{equation}
In the limit of vanishing quark masses the chiral condensate stays
non zero as long as chiral symmetry is spontaneously broken. The 
chiral condensate thus is an obvious order parameter in the chiral
limit,  
\begin{equation}
\langle \bar{\chi}\chi \rangle \; \cases{
>\; 0 \; \Leftrightarrow \;{\rm symmetry~broken~ phase,}& $T\; <\; T_c$ \cr
=\; 0 \; \Leftrightarrow \;{\rm symmetric~ phase,}&
$T\; > \; T_c$} \quad .
\label{chiorder}
\end{equation}

For light quarks the global chiral symmetry is expected to control the 
critical behaviour of the QCD phase transition. In particular, the order
of the transition is expected to depend on the number of light or
massless flavours. The basic aspects of the
$n_f$-dependence of the phase diagram have been derived by Pisarski and
Wilczek \cite{Pis84} from an effective, 3-dimensional  Lagrangian for
the order parameter\footnote{It should be noted that this ansatz
assumes that chiral symmetry is broken at low temperatures. Instanton
model calculations suggest that the vacuum, in fact, is chirally
symmetric already for $n_f\ge 5$ \cite{Shu95}.},
\begin{eqnarray}
\label{effchiralL}
{\cal L}_{eff} =&~& -\frac{1}{2} {\rm Tr}(\partial_\mu 
\Phi^\dagger \partial^\mu \Phi)
  - \frac{1}{2} m^2 {\rm Tr}(\Phi^\dagger \Phi) + \frac{\pi^2}{3} g_1
\left({\rm Tr}(\Phi^\dagger \Phi)\right)^2 \nonumber \\
&~&  + \frac{\pi^2}{3} g_2 {\rm Tr}\left( (\Phi^\dagger \Phi)^2\right) + c
\left(\det \Phi + \det \Phi^\dagger \right)
\quad ,
\end{eqnarray}
with $\Phi \equiv (\Phi_{ij})$, $i,j=1,..., n_f$. ${\cal L}_{eff}$ has 
the same global symmetry as the QCD Lagrangian. A renormalization group 
analysis of this Lagrangian suggests that the transition is 
first order for $n_f \ge 3$ and second order for $n_f =2$. The
latter, however, is expected to hold only if the axial $U_A(1)$ symmetry
breaking, related to the $\det \Phi$ terms in Eq.~\ref{effchiralL}, 
does not become too weak at $T_c$ so that the occurrence of a fluctuation 
induced first order transition would also become possible. 

This basic pattern has indeed been observed in lattice calculations. 
So far no indication for a discontinuous transition has been observed
for $n_f =2$. The transition is found to 
be first order for $n_f \ge 3$. Moreover, the transition temperature 
is decreasing
with increasing $n_f$ and there are indications that chiral symmetry
is already restored in the vacuum above a critical number of 
flavours \cite{Iwa94}. 

\begin{figure}[htb]
\begin{center}
\epsfig{bbllx=0,bblly=175,bburx=564,bbury=514,
file=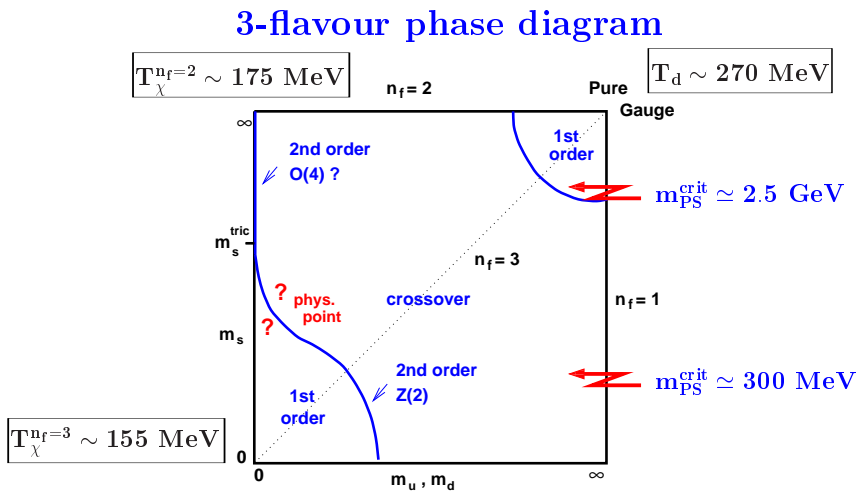,width=110mm}
\end{center}
\caption{The QCD phase diagram of 3-flavour QCD with degenerate
(u,d)-quark masses and a strange quark mass $m_s$. }
\label{fig:phased}
\end{figure}

The anticipated phase diagram of 3-flavour QCD at vanishing baryon number 
density is shown in Fig.~\ref{fig:phased}.
An interesting aspect of the phase diagram is the occurrence of 
a second order transition line in the light quark mass regime, 
the boundary of the region of first order phase transitions.
On this line the transition is controlled by an effective 
3-dimensional theory with global $Z(2)$ symmetry \cite{Gav94},
which is not a symmetry of the QCD Lagrangian. As this boundary
lies in the light quark mass regime it may well be that this
second order transition, for which neither the chiral condensate 
nor the Polyakov loop will be the order parameter,
is equally important for the critical or crossover behaviour 
of QCD with a realistic quark mass spectrum as the nearby critical
point in the chiral limit. In particular, we note that the critical
exponent $\alpha$ is positive for the 3-$d$, $Z(2)$ symmetric models
whereas it is negative for the $O(4)$ model. A nearby $Z(2)$ symmetric
critical point in the QCD phase diagram will thus induce larger density 
fluctuations than would be expected in the vicinity of the
chiral critical point. It therefore will be important to determine in 
detail the location of the physical point in the QCD phase diagram.

\section{Deconfinement versus Chiral Symmetry Restoration}

As outlined in the previous section the two properties of QCD, which
explain the basic features of the observed spectrum of hadrons,
are also of central importance for the structure of the
QCD phase diagram at finite temperature -- confinement and chiral 
symmetry breaking.
While the former explains why we observe only colourless states in the
spectrum the latter describes the presence of light Goldstone
particles, the pions. The confining property of QCD manifests itself 
in the long range behaviour of the heavy quark
potential. At zero temperature the potential rises linearly at large
distances\footnote{Here large distances actually refer to $r\simeq 1$~fm. 
For larger distances the
spontaneous creation of quark anti-quark pairs from the vacuum leads to
a breaking of the string, {\it i.e.} the potential tends to a constant
value for $r \rightarrow \infty$ (see Fig.~4).},
$V_{\bar{q}q} (r) \sim \sigma r$, where $\sigma \simeq (425~{\rm MeV})^2$ 
denotes the string tension, and forces the quarks and gluons
to be confined to a {\it hadronic bag}. Chiral symmetry breaking leads
to a
non-vanishing quark anti-quark condensate,
$\langle \bar{q}q  \rangle \simeq (250~{\rm MeV})^3$ in the vacuum.
Inside the hadron bag, however, the condensate vanishes.
At high temperatures the individual hadronic bags are expected to merge
to a single large bag, in which quarks and gluons can move freely. 
This bag picture is closely related to percolation models for the
QCD phase transition \cite{Satz}. It provides an intuitive argument for 
the occurrence of deconfinement and chiral symmetry restoration.
A priory it is, however, not evident that both
non-perturbative properties have to get lost at the same temperature.
It has been speculated that two distinct phase transitions leading
to deconfinement at $T_d$ and chiral symmetry restoration at $T_\chi$
could occur in QCD \cite{Shuryak}. General arguments about the scales
involved\footnote{The hadronic bag is larger than the constituent quark 
bag of a current quark surrounded by its gluon cloud.}  
suggest that $T_d \le T_\chi$. Two distinct phase transitions
indeed have been found in QCD related models like the $SU(3)$ gauge 
theory with adjoint fermions \cite{Kar99a}. In QCD, however, there
seems to be only one transition from the low temperature hadronic
regime to the high temperature plasma phase. In fact, as can be
seen from Fig.~\ref{fig:phased} there is a wide range of parameters
(quark masses) for which the transition is not related to any
singular behaviour in thermodynamic observables; instead of  
a phase transition one observes just a rapid crossover behaviour.
It thus is legitimate to ask which 
thermodynamic properties change when one moves from the low to the high 
temperature regime and to what extent these changes are related to 
deconfinement and/or chiral symmetry restoration. 

In the previous section we have introduced order parameters
for deconfinement in the infinite quark mass limit, $\langle L \rangle$,
and chiral symmetry restoration in the limit of vanishing
quark masses, $\langle \bar{\psi}\psi \rangle$. 
Related observables, which also signal a sudden change in the long
distance behaviour of the heavy quark potential or the chiral 
condensate as function of temperature, are the corresponding 
susceptibilities, the Polyakov loop susceptibility ($\chi_L$) and 
the chiral susceptibility ($\chi_{m}$),
\begin{equation}
\chi_L = N_\sigma^3\biggl(\langle L^2 \rangle - \langle L \rangle^2\biggr)~~
\quad, \quad
\chi_{m} = {\partial \over \partial m_q} \langle \bar{\psi}\psi
\rangle
\quad .
\label{sus}
\end{equation}

\begin{figure}[t]
\begin{center}
\epsfig{bbllx=105,bblly=220,bburx=460,bbury=595,clip=,
file=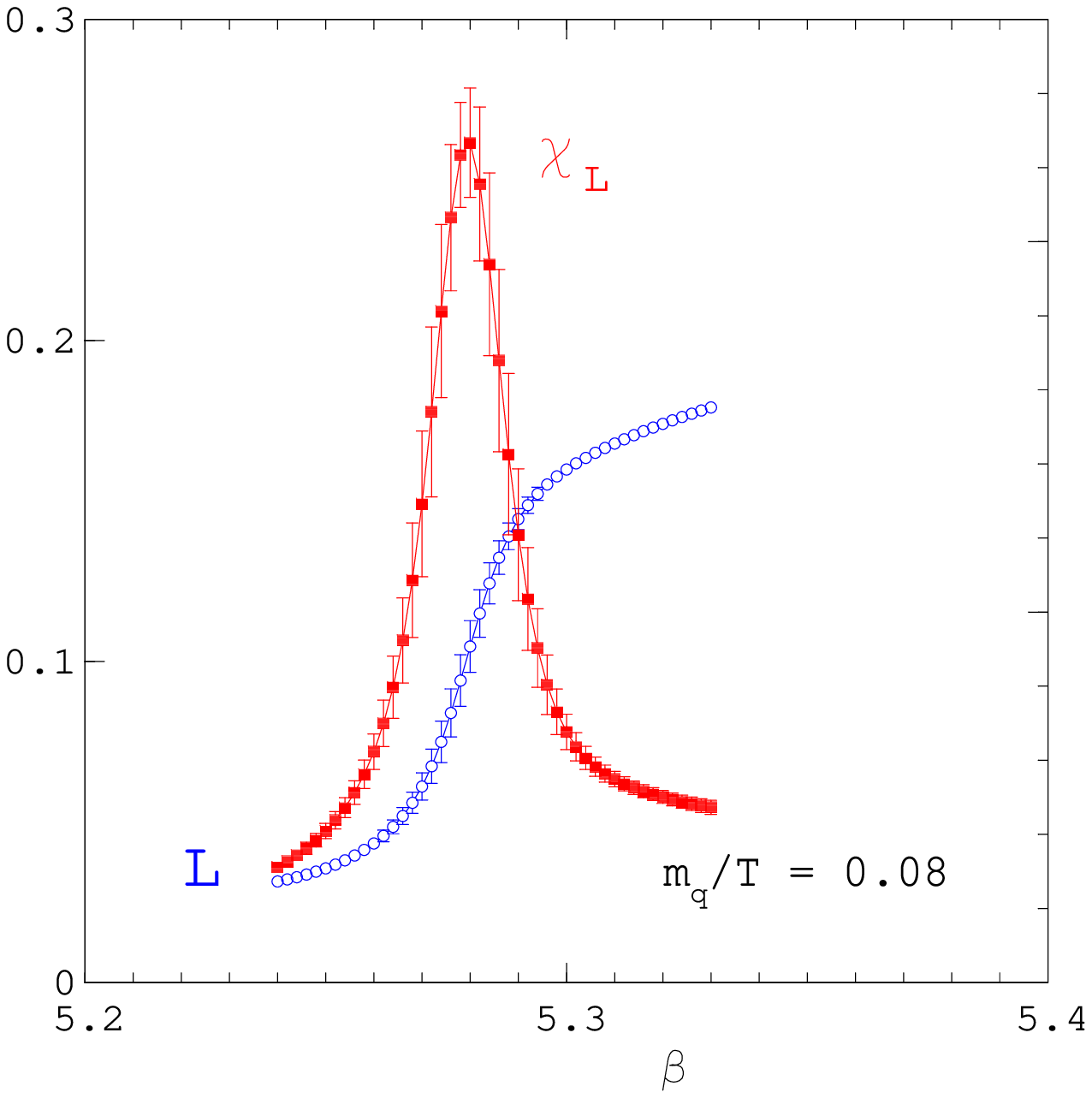,height=60mm}
\epsfig{bbllx=105,bblly=220,bburx=460,bbury=595,clip=,
file=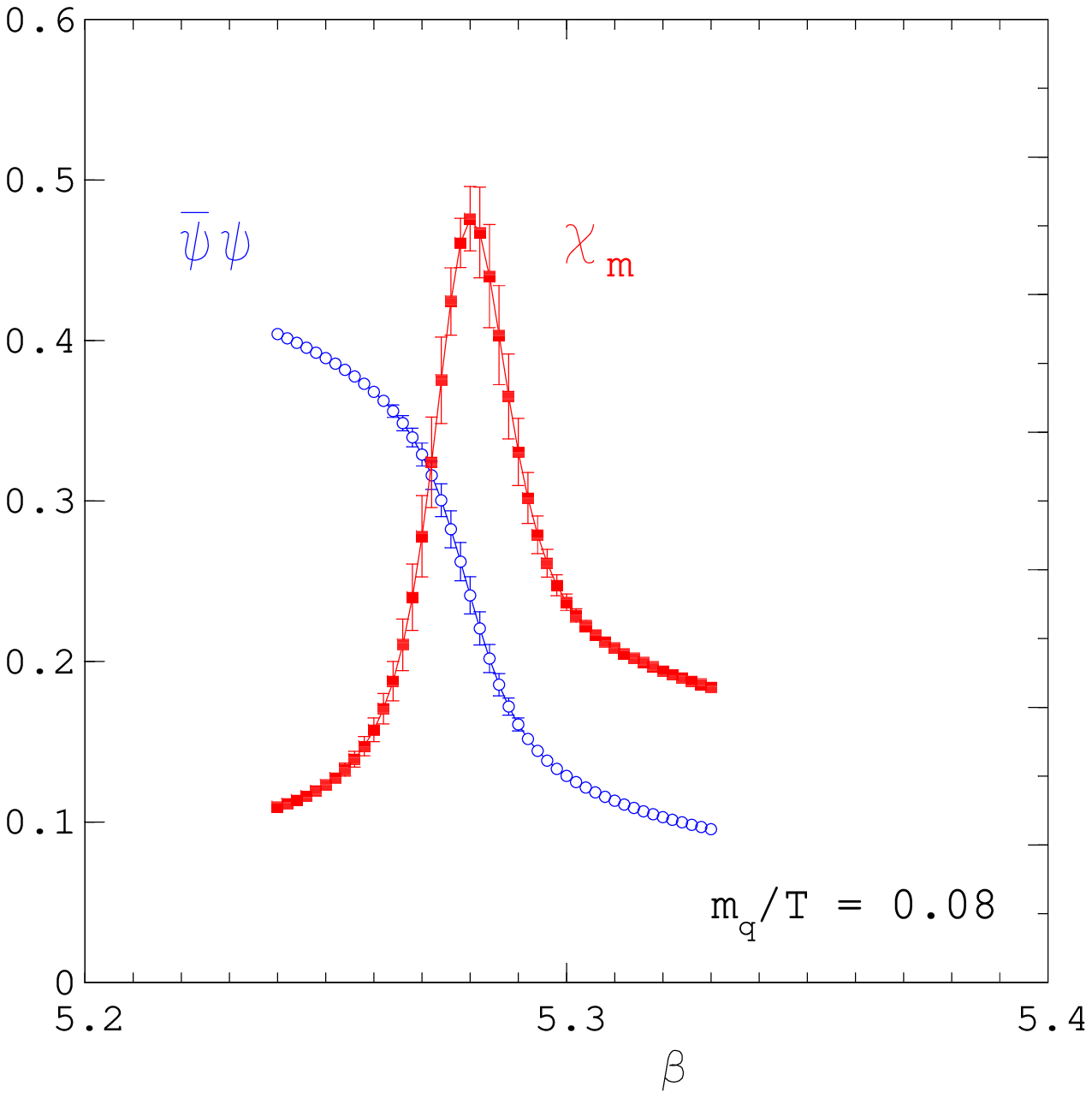,height=60mm}
\end{center}
\caption{Deconfinement and chiral symmetry restoration in 2-flavour QCD:
Shown is $\langle L\rangle$ (left), which is the order parameter for 
deconfinement in the pure gauge limit ($m_q\rightarrow \infty$), and 
$\langle \bar{\psi}\psi \rangle$ (right), which is the order parameter
for chiral symmetry breaking in the chiral limit ($m_q\rightarrow 0$).  
Also shown are the corresponding susceptibilities as a function of the 
coupling $\beta=6/g^2$.}
\label{fig:demo}
\end{figure}

\begin{figure}[t]
\begin{center}
\epsfig{file=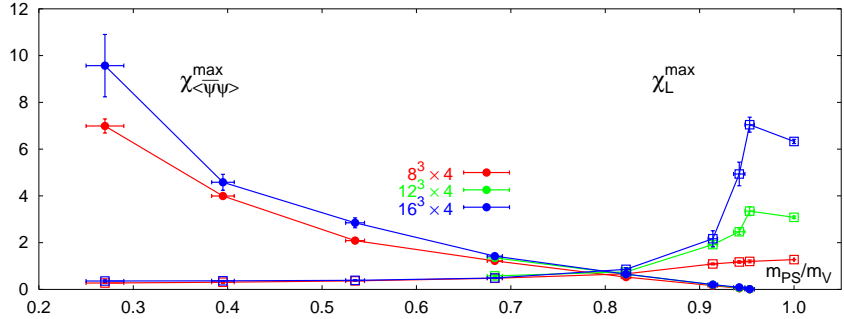,
width=115mm}
\end{center}
\caption{Quark mass dependence of the Polyakov loop and chiral
susceptibilities versus $m_{PS}/m_V$ for 3-flavour QCD. Shown are
results from calculations with the improved gauge and staggered 
fermion action discussed in the Appendix.}
\label{fig:chi_L_sus}
\end{figure}

The behaviour of these observables is shown in Fig.~\ref{fig:demo}
for the case of two flavour QCD with light quarks. This clearly
shows that the gauge coupling at which the different susceptibilities
attain their maxima, or correspondingly the points of most
rapid change in $\langle L\rangle$ and $\langle \bar{\psi}\psi \rangle$
coincide.
Calculations of these observables for QCD with three degenerate quark 
flavours\footnote{This corresponds to calculations along the dotted,
diagonal line in the phase diagram shown in Fig.~\ref{fig:phased}.} 
have been performed for a wide range of quark masses \cite{Kar01}. 
They confirm that the location of maxima in both susceptibilities are 
indeed strongly correlated. Within statistical accuracy they occur at the 
same temperature, although the height of these maxima is strongly 
quark mass dependent. This is shown in Fig.~\ref{fig:chi_L_sus}. 
For large ($m_{PS}/m_V \gsim 0.9$) and small ($m_{PS}/m_V \lsim 0.3$) 
quark masses the Polyakov loop and chiral susceptibility, respectively, 
show a strong volume dependence, which is indicative for the presence of
first order phase transitions in these corners of the phase diagram. 
Using zero temperature string tension calculations the pseudo-scalar
meson masses have been estimated at which the first order transitions
end in a second order transition. For better orientation in the phase 
diagram these estimates,
which at present are not well established and certainly are still
subject to lattice artifacts (discretization errors, flavour symmetry
breaking), are shown in Fig.~\ref{fig:phased}. 
As can be seen there is a broad range of quark (or
meson) masses for which the QCD transition to the high temperature
phase is a non-singular crossover.  

As expected the first order phase transition in the large quark
mass regime is most clearly visible in the behaviour of the Polyakov 
loop susceptibility, {\it i.e.} the fluctuation of the order parameter
for the confinement-deconfinement transition in the pure gauge
($m_q \rightarrow \infty$) limit. Similarly, the transition in
the chiral limit is most pronounced in the behaviour
of the chiral condensate and its susceptibility. 
This emphasizes the chiral aspects of the QCD transition.
One thus may wonder in what respect this transition in the light quark 
mass regime is a deconfining transition.

\subsection{Deconfinement}

\begin{figure}[t]
\begin{center}
\hspace*{-0.4cm}
\epsfig{file=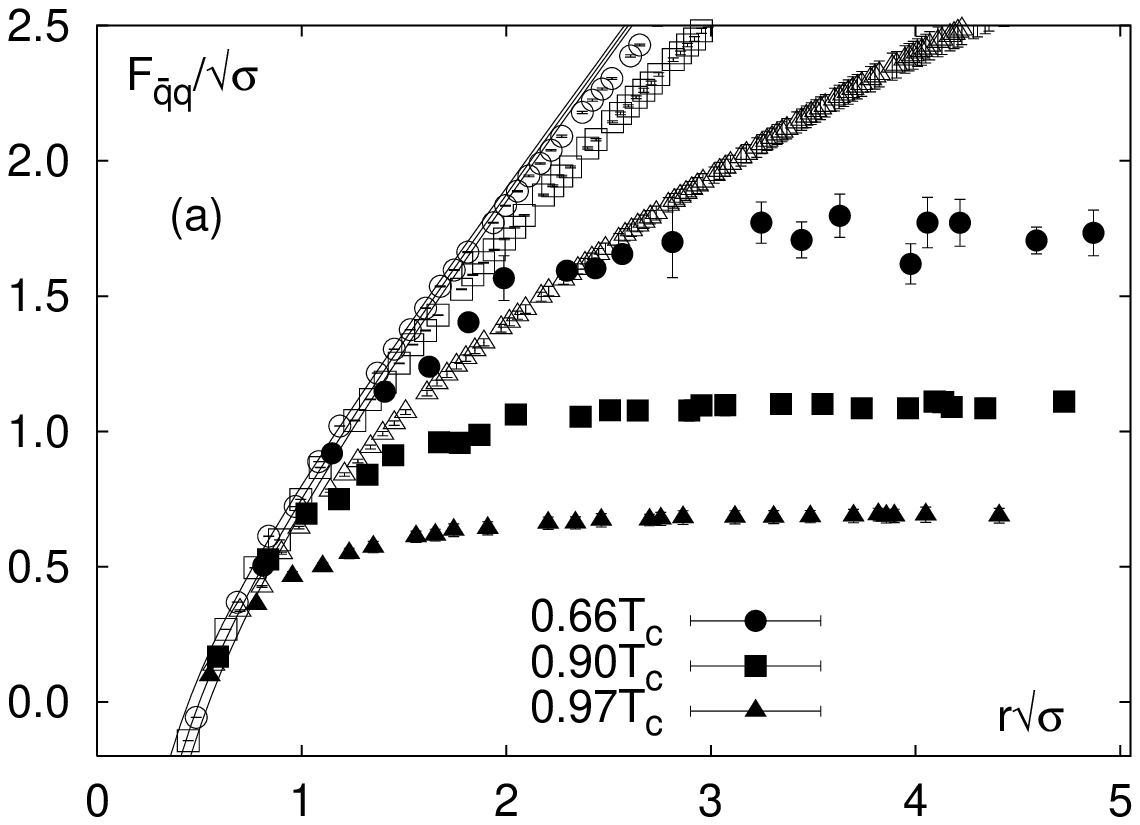,
width=62mm}\hspace*{-0.3cm}
\epsfig{file=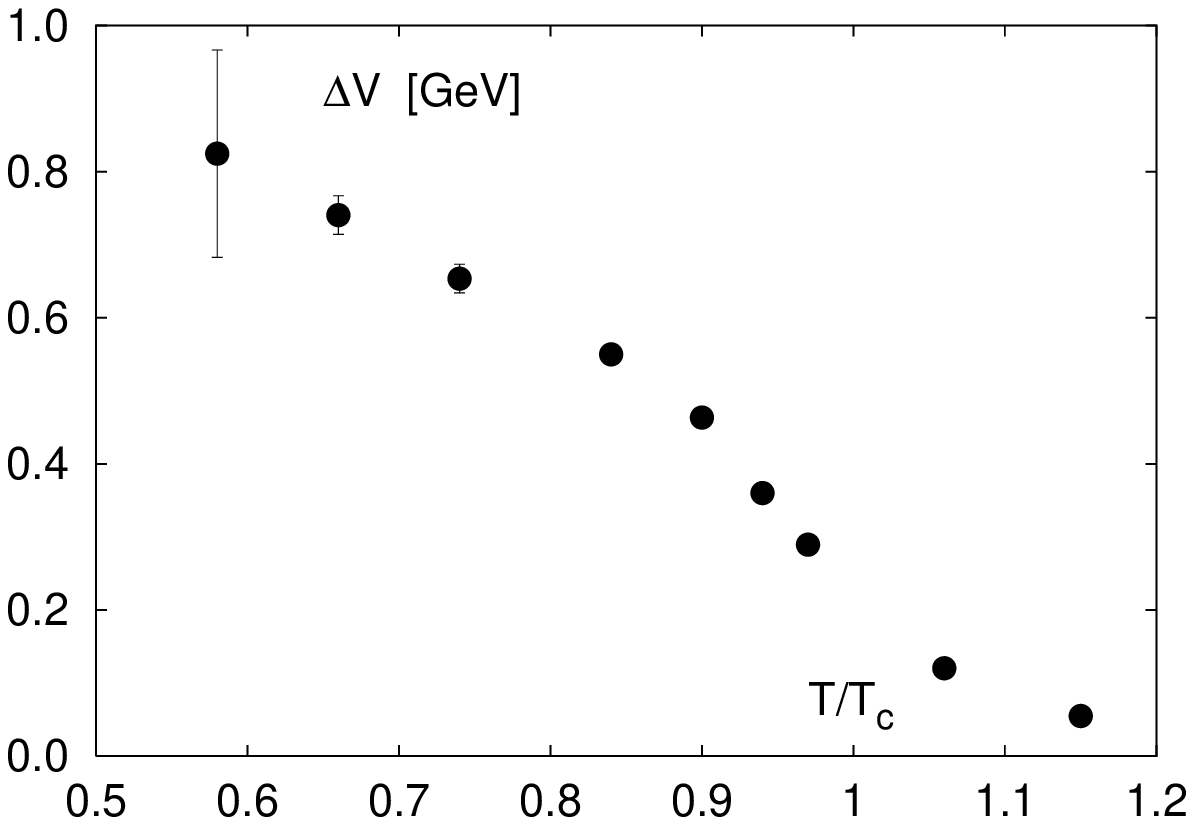,
width=62mm}
\end{center}
\caption{The left hand figure shows the heavy quark free energy in units 
of the square root of the string tension for the SU(3) gauge theory 
(open symbols) and three flavour QCD with light quarks (full symbols). 
The right hand figure gives the limiting value of the free energy
normalized to the value at distance $r\simeq 0.23$~fm \cite{Kar01}
as a function of temperature for the case of three flavour QCD.
The quark mass used in the $n_f=3$ calculations corresponds to a ratio of 
pseudo-scalar and vector meson masses of $m_{PS} / m_V\simeq 0.7$.}
\label{fig:pot_pg_nf3}
\end{figure}

When talking about deconfinement in QCD we have in mind that a large number
of new degrees of freedom gets liberated at a (phase) transition 
temperature; quarks and gluons which at low temperature are confined 
in colourless hadrons and thus do not contribute to the thermodynamics, 
suddenly become liberated and start contributing to bulk thermodynamic 
observables like the energy density or pressure. In the heavy quark mass
limit ($m_q \equiv \infty$) a more rigorous statement is based 
on the analysis of the long distance behaviour of the heavy quark free 
energy, which approaches a constant above $T_c$ but diverges 
for $T < T_c$. For finite quark masses, however, the long
distance behaviour of $F_{\bar{q}q}$ can no longer serve as an order
parameter, the heavy quark free energy stays finite for all 
temperatures. This is shown in Fig.~\ref{fig:pot_pg_nf3} where
we compare results from calculations in the pure gauge theory \cite{Kac00}
with results from a calculation in three flavour QCD \cite{Kar01}.
It is apparent from this figure that there is no
drastic qualitative change in the structure of the heavy quark free 
energy as one crosses the transition temperature. Although the 
most rapid change in 
$\Delta V \equiv F_{\bar{q}q}(\infty) -
F_{\bar{q}q}(r=0.5/\sqrt{\sigma})$ occurs for  $T\simeq T_c$
and thus resembles the behaviour of the Polyakov loop expectation
value and its susceptibility shown in Fig.~\ref{fig:demo}, 
the heavy quark free energy does not seem to be a good indicator
for deconfinement in the presence of light quarks. It looses its
role as a rigorous order parameter and also does not reflect changes 
in the number of partonic degrees of freedom contributing to the 
thermodynamics. 

On the other hand, we expect that bulk thermodynamic
quantities like the pressure do reflect the relevant number
of degrees of freedom contributing to the thermodynamics  
in the high temperature limit. Due
to asymptotic freedom the QCD pressure will approach the ideal
gas value at infinite temperature. In this limit the number
of degrees of freedom (quarks+gluons) is much larger than the
three light pions which dominate the thermodynamics at low temperature,

\begin{equation}
{p \over T^4} = \cases{3 {\pi^2 \over 90} &,~ $T \rightarrow 0$ \cr 
 (16 + \frac{21}{2} n_f) {\pi^2 \over 90} &,~
$T \rightarrow \infty$} \quad . 
\label{pSB}
\end{equation}

\begin{figure}[t]
\begin{center}
\epsfig{
file=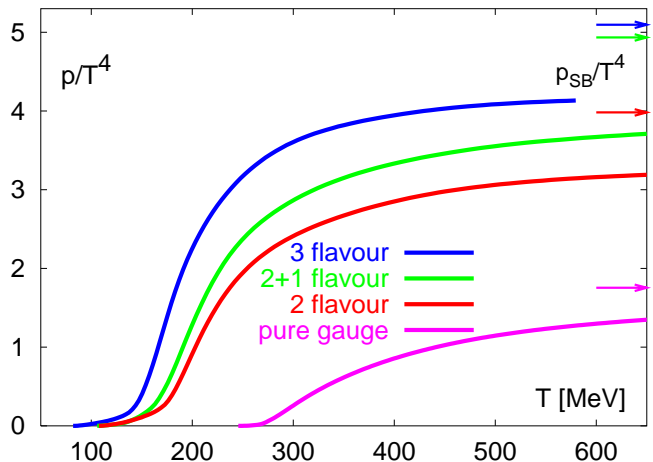,
width=95mm}
\end{center}
\caption{The pressure in QCD with different number of degrees of 
freedom as a function of temperature. The curve labeled (2+1)-flavour
corresponds to a calculation with two light and a four times heavier
strange quark mass \cite{Kar00a}.} 
\label{fig:pressure}
\end{figure}

This change of active degrees of freedom is clearly
visible in calculations of e.g. the pressure 
in the pure gauge sector and for QCD with different numbers of 
flavours. As can bee seen in Fig.~\ref{fig:pressure} 
the pressure strongly reacts to
changes in the number of degrees of freedom.  
It is this drastic change in the behaviour
of the pressure or the energy density which indicates that
the QCD (phase) transition to the plasma phase indeed is 
deconfining. However, it also is worthwhile to note 
that the transition does, in fact, take place at rather small
values of the pressure (and energy density). 
Only for temperatures $T\gsim 2T_c$ does the pressure come 
close to the ideal gas limit so that one can, with some 
justification, identify the corresponding light degrees of freedom.
This is the case
for QCD with light quarks as well as in the quenched limit.
At least for temperatures up to a few times $T_c$ the dynamical 
degrees of freedom are certainly not just weakly interacting 
partons.

\subsection{Chiral symmetry restoration}

\begin{figure}[htb]
\begin{center}
\epsfig{bbllx=127,bblly=264,bburx=451,bbury=587,
file=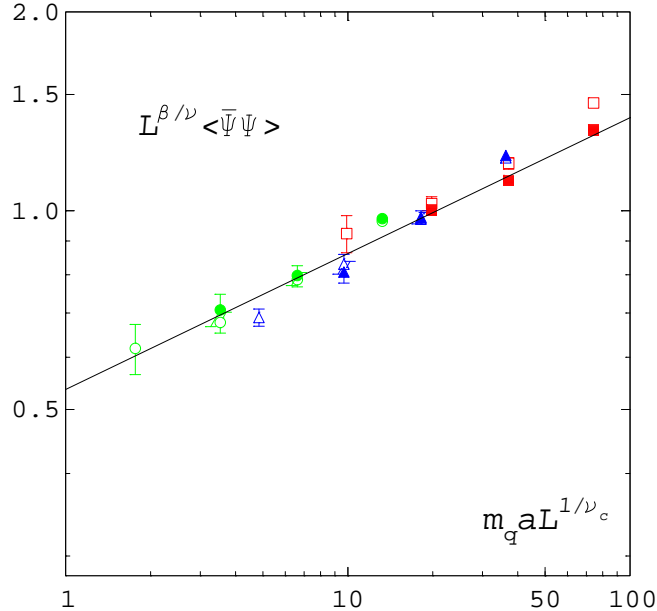,width=75mm}
\end{center}
\vspace*{0.5cm}
\caption{Finite size scaling of the chiral condensate in 2-flavour
QCD \cite{Eng01b}. Shown are data from calculations with standard staggered 
fermions on lattices of size $8^3\times 4$ (circles), $12^3\times 4$ 
(triangles) and $16^3\times 4$ (squares). The calculations have been 
performed with different values of the quark mass  at fixed value
of the scaling variable $z_c$ corresponding to the pseudo-critical
line (Eq.~\ref{pcline}).}
\label{fig:qcdfsc}
\end{figure}

As chiral symmetry restoration does not lead to a significant
change of light degrees of freedom, it also is not expected to have an
appreciable effect on bulk thermodynamic observables -- apart from 
controlling details of the transition very close to $T_c$. In particular,
we expect that in the case of a continuous transition for $n_f=2$,
the chiral order parameter and its derivative, the chiral susceptibility,
show critical behaviour which is characteristic for $O(4)$ spin models
in three dimensions \cite{Pis84}. The expected critical behaviour
follows from standard scaling arguments derived from the singular 
part of the free energy density,
\begin{equation}   
f_s(t,h)\equiv - \frac{T }{ V} \ln Z_s   = b^{-d}f_s(b^{y_t}t, b^{y_h} h)
\quad ,   
\label{singularfree} 
\end{equation} 
where $t=(T-T_c)/T_c \sim (\beta -\beta_c)$ is the reduced temperature,
$h=m/T \sim m_qN_\tau$ the scaled quark mass and $b$ is an arbitrary
scale factor.  For the chiral order
parameter, $\langle \bar{\psi} \psi \rangle$, and 
the chiral susceptibility, $\chi_m$, one finds from Eq.~\ref{singularfree},
\begin{eqnarray}
\langle \bar{\psi} \psi \rangle
    &=&  h^{1/\delta} F(z) \label{mageos} \\
    \chi_m (t,h) &=&  \frac{1}{\delta} h^{1/\delta - 1}
    \biggl[ F(z) - \frac{z}{\beta} F'(z) \biggr]
    \quad ,
\label{scale}
\end{eqnarray}
with scaling functions $F$ and $F'$ that only depend on a
specific combination of the reduced temperature  and scaled
quark mass, $z=th^{-1/ \beta\delta}$. The critical
exponents $\beta$ and $\delta$ are
given in terms of $y_t$ and $y_h$ as  $\beta = (1-y_h)/y_t$ and
$\delta = y_h/(1-y_h)$. As the $t$-dependence enters in $\chi_m (t,h)$ 
only through $z$ one also deduces that the line of pseudo-critical
couplings defined through the location of the maximum of $\chi_m (t,h)$
at fixed $h$ is described by a universal scaling function, 
\begin{equation}
\label{pcline}
t_\mathrm{c}(h)\equiv z_\mathrm{c}\; h^{1/ \beta\delta} \quad . 
\end{equation}
Although there is ample evidence that the phase transition
in 2-flavour QCD is continuous in the chiral limit, the evidence
for the expected $O(4)$ scaling is, at present, ambiguous.
The behaviour of the pseudo-critical couplings is, in general,
consistent with the expected scaling behaviour. The information
on the magnetic equation of state, Eq.~\ref{mageos}, however,
seems to depend on the fermion discretization scheme used to 
analyze the critical behaviour.
While calculations with Wilson fermions yield
almost perfect agreement with the universal form of the $O(4)$ 
magnetic equation of state \cite{Yoshie}, significant deviations have 
been found in the case of staggered fermions \cite{MilcO4}.
The failure of the scaling analysis in the case of staggered
fermions is surprising as this staggered fermion action has a global 
$O(2)$ symmetry even for finite values of the lattice cut-off and as the 
$O(4)$ and $O(2)$ magnetic equations of state are quite similar
\cite{Eng01a}. this suggests that finite size effects still play an
important role, which is supported by a recent finite size scaling
analysis \cite{Eng01b}. 
In Fig.~\ref{fig:qcdfsc} we show the finite size scaling behaviour of 
the chiral condensate, which has been reanalyzed in \cite{Eng01b}.  
It is consistent with $O(4)$ (or $O(2)$) scaling behaviour.  
This aspect, however, clearly needs further studies. 

The changes of the chiral condensate below $T_c$ and chiral symmetry 
restoration at $T_c$ will have a strong influence on the 
light hadron spectrum. At $T_c$ the pseudo-scalar mesons (pions) will 
no longer be Goldstone particles, they turn into massive modes (quasi-particle
excitations?) above $T_c$. Long distance correlations of the chiral 
condensate decay exponentially with a characteristic length scale
proportional to the inverse scalar meson mass. A diverging
chiral susceptibility at $T_c$ thus indicates that the scalar meson 
mass vanishes at $T_c$.  
The mass splitting between parity partners thus will decrease when
the symmetry breaking reduces and finally will become degenerate
at $T_c$. 

\begin{figure}[htb]
\begin{center}
\epsfig{
file=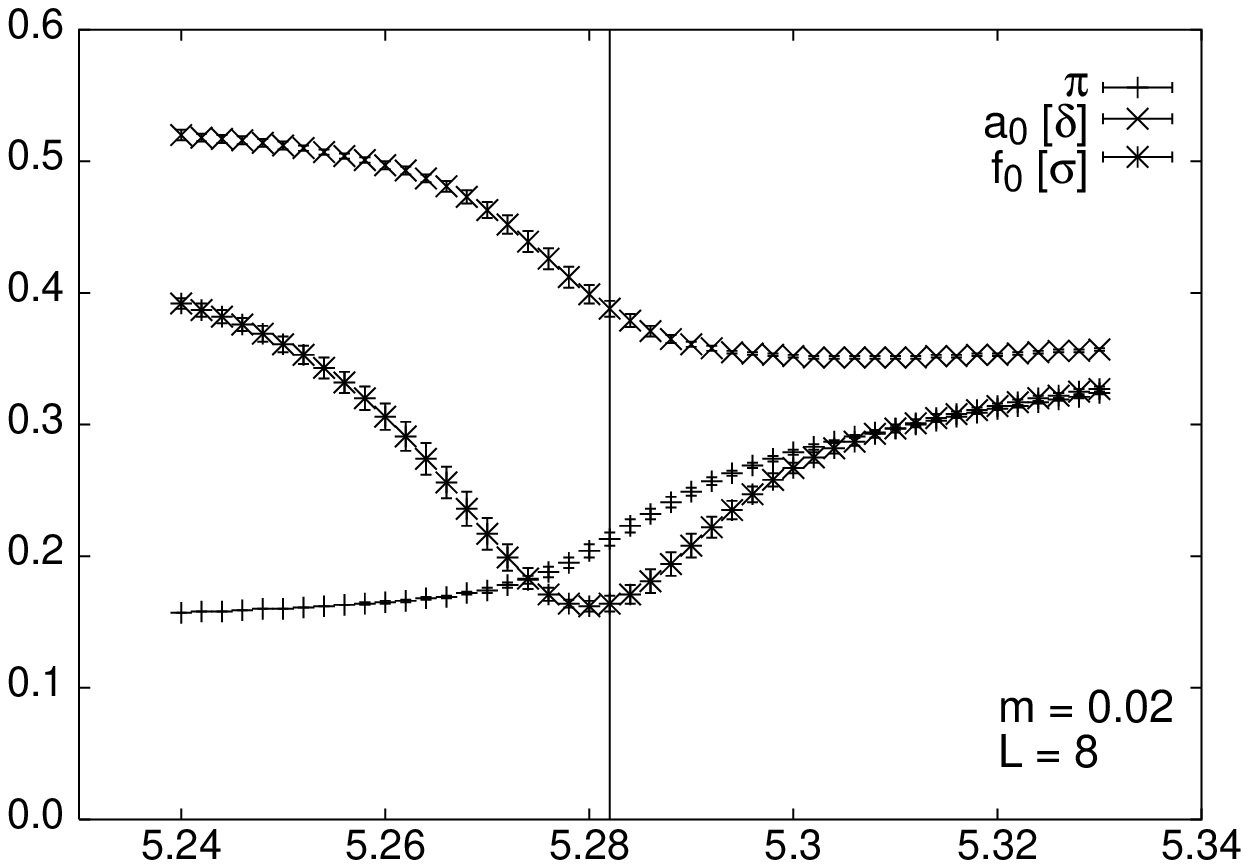,
width=95mm}
\end{center}
\caption{Temperature dependence of generalized hadron masses extracted 
from hadronic susceptibilities. Shown are results from calculations
in 2-flavour QCD performed on lattices of size $8^3\times 4$ with 
staggered fermions of mass $m_q=0.02$.} 
\label{fig:hadrons}
\end{figure} 

As indicated above the modifications of the hadron spectrum are 
reflected by the temperature dependence of appropriately chosen
susceptibilities, which are the space-time integral over hadronic
correlation functions in a given quantum number channel,  
\begin{equation}
\chi_H = \int_0^{1/T} {\rm d}\tau \int {\rm d}^3r \; G_H (\tau,\vec{r})
\quad ,
\label{hadronic}
\end{equation}
where the hadronic correlation function $G_H (\tau,\vec{r})$ for 
mesons is given by,
\begin{equation}
G_H (\tau,\vec{r}) = \langle \bar{\chi}(0)\Gamma_H \chi (0)
\bar{\chi}(\tau,\vec{r})\Gamma_H \chi (\tau,\vec{r}) \rangle \quad ,
\label{correlator}
\end{equation}
and $\Gamma_H$ is an appropriate combination of $\gamma$-matrices that
projects onto a chosen quantum number channel. In particular,
we note that the chiral susceptibility, $\chi_m$, defined in 
Eq.~\ref{sus} is the susceptibility of the scalar correlation function.
These susceptibilities define generalized masses, $m_H^{-2} \equiv \chi_H$, 
which are shown in Fig.~\ref{fig:hadrons}. They, indeed, show
the expected behaviour; scalar ($f_0$) and pseudo-scalar ($\pi$) 
partners become degenerate at $T_c$ whereas the vector meson ($\delta$)
which is related to the scalar meson through a $U_A(1)$ rotation
only gradually approaches the other masses. The axial $U_A(1)$
symmetry thus remains broken at $T_c$.

\section{Screening at High Temperature -- Short versus Long Distance Physics}

Our picture of the thermodynamics in the high temperature phase 
of QCD largely is influenced by perturbative concepts -- asymptotic
freedom and the screening of electric and magnetic components of
the gluon fields. Asymptotic freedom suggests that the temperature 
dependent running coupling,
$g(T)$, becomes small at high temperatures and eventually vanishes
in the limit $T\rightarrow \infty$. This in turn will
lead to a separation of the thermal length scale, $1/T$, from 
the electric, $1/g(T)T$, and magnetic, $1/g^2(T)T$, screening length 
scales. The experience gained from lattice calculations in the
pure $SU(3)$ gauge theory, however, suggests that this separation of 
scales, unfortunately, will set in only at asymptotically
large temperatures. For all interesting temperatures reachable in heavy
ion experiments or even covering the temperature interval between
the strong and electroweak phase transitions that occurred in the
early universe, the coupling $g(T)$ is of ${\cal O}(1)$ and, moreover, 
the Debye screening mass is significantly larger than the leading order 
perturbative value,
\begin{equation}
m_\mathrm{D} = \sqrt{1  +\frac{n_\mathrm{f}}{6} }\;
g(T)\: T\quad . 
\label{debye}
\end{equation}
In fact, for $T_c \lsim T \lsim 100 \; T_c$ one 
finds that $m_D$ is still about three times larger than this leading 
order value \cite{Kaj97}. A consequence of this large
value of the screening mass is that also short distance properties of 
the plasma are strongly influenced by non-perturbative screening 
effects; the Debye screening length, $r_D \equiv 1/m_D$, becomes
compatible with the characteristic length scale $r_{SB}$ in a free 
gas where the main contribution to the Stefan-Boltzmann law originates 
from particle with momenta $p \sim 3T$, {\it i.e.} $r_{SB} \sim 1/3T$,
and is of the same order as the mean separation between partons
in a quark-gluon plasma. 
We thus must expect that non-perturbative screening effects also have
an influence on bulk thermodynamics properties (pressure,
energy density) above $T_c$ and even up to quite large temperatures. 
In fact, the calculations of the pressure and energy density in the 
$SU(3)$ gauge theory and in QCD with light quarks, which we are
going to discuss in the next section, show that at temperatures a 
few times $T_c$ deviations from the ideal gas limit are still too large 
to be understood in terms of conventional high temperature perturbation 
theory, which converges badly at these temperatures just because of
the large contribution arising from Debye screening \cite{Zhai}.

\begin{figure}[htb]
\begin{center}
\epsfig{file=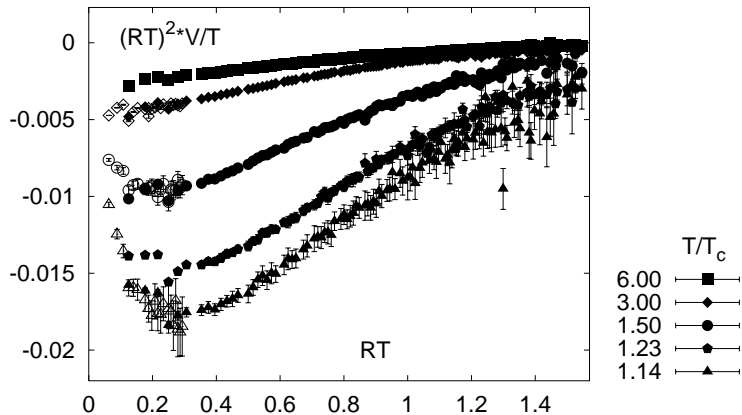,width=85mm}
\end{center}
\caption{The heavy quark free energy at various temperatures in the 
deconfined phase of the $SU(3)$ gauge theory. Calculations have been 
performed on lattices of size $32^3 \times 8$ (filled symbols) and 
$32^3\times 16$ (open symbols) \cite{Zantow}.}
\label{fig:screen} 
\end{figure}

The screening of static quark and anti-quark sources is commonly
analyzed in terms of Polyakov-loop correlation functions, 
which define the heavy quark free energy introduced in Eq.~\ref{poly}.
The leading perturbative contribution to $F_{\bar{q}q}(r,T)$
results from the exchange of two gluons, 
\begin{eqnarray}
{V_{\bar{q}q}(r,T) \over T} \equiv
{F_{\bar{q}q}(r,T) - F_{\bar{q}q}^\infty \over T} &=&
-  \ln\biggl(
{\langle {\rm Tr} L_{\vec{x}} {\rm Tr} L^{\dagger}_{\vec{y}} \rangle
\over |\langle L \rangle |^2} \biggr) \nonumber \\
&=& - {1\over 16}
\biggl({g^2(T) \over 3\pi } {1 \over rT}\biggr)^2 + {\cal O}(g^5) \quad .
\label{polyasy}
\end{eqnarray}
Higher order
contributions will lead to screening of this powerlike large
distance behaviour, $1/rT \;\rightarrow \; \exp{(-m_D r)}/rT$, and
also result in an exponentiation of the leading order contribution.
We then may split $F_{\bar{q}q}$ in contributions arising from 
quark anti-quark pairs in singlet ($F_1$) and octet ($F_8$) 
configurations \cite{McLerran},
\begin{equation}
\E^{-F_{\bar{q}q}(r,T)/T } =
\frac{1}{9} \;  \E^{- F_{1}(r,T)/T } +
\frac{8}{9}   \; \E^{- F_{8}(r,T)/T }   
\quad .
\label{average}
\end{equation}     
In accordance with zero temperature perturbation theory
the singlet free energy is attractive whereas the octet free energy 
is repulsive. Their relative strength is such that it leads to a 
cancellation of the leading $\mathcal{O}(g^2)$ contributions to 
the colour averaged heavy quark free energy $F_{\bar{q}q}$. From 
Eq.~\ref{average} it is, however, apparent, that the cancellation of 
singlet and octet contributions only occurs at large distances.
At short distances the contribution from the attractive singlet 
channel will dominate the heavy quark free energy,
\begin{eqnarray}
{F_{\bar{q}q}(r,T) \over T} &=& {F_{1}(r,T) \over T}\; +\; {\rm const.}
\nonumber \\
&=& - {g^2(T) \over 3\pi}{1\over rT} \; +\; {\rm const.} 
\quad\quad {\rm for}\quad \; rT << 1 \quad . 
\label{short}
\end{eqnarray}
In order to eliminate the subleading power-like behaviour at large
distances we show in  Fig.~\ref{fig:screen} 
$(rT)^2\; V_{\bar{q}q}(r,T)/T$ calculated for the $SU(3)$ gauge 
theory. As can be seen the change from the Coulomb-like behaviour at
short distances to the exponential screening at large distances 
can be well localized. For
$T_c \le T \lsim 2T_c$ it occurs already for $rT \simeq 0.2$ or 
$r \simeq 0.15\; (T_c/T)\; {\rm fm}$ and  
shifts slightly to smaller $rT$ with increasing temperature. 
A consequence of this 
efficient screening at short distances is that even heavy quark
bound states get destroyed close to $T_c$ in the plasma
phase ($J/\psi$-suppression \cite{Matsui}). 

The perturbative analysis of the heavy quark free energy also
suggests that for fixed $rT$ the only temperature dependence
of $V(r,T)/T$ arises from the running of the coupling $g(T)$.
The rapid change of $V(r,T)/T$ at fixed $rT$ which is apparent 
in  Fig.~\ref{fig:screen} thus also suggests that for 
temperatures $T\lsim 3T_c$ the coupling $g(T)$ varies much
more rapidly than the asymptotically expected logarithmic 
running with $T$. 

We thus conclude that non-perturbative screening
effects are important for the thermodynamics in the plasma phase
also for short distance observables which are sensitive to the 
physics at distances $r\gsim 1/5T$. The strong temperature 
dependence observed for $T\lsim 3T_c$, moreover, suggests 
that the system cannot be described at a weakly coupled, 
asymptotically free plasma at these temperatures.  These general 
features will carry over to the temperature dependence of the QCD 
equation of state which we are going to discuss in the next section.

\section{The QCD Equation of State}

The most fundamental quantity in equilibrium thermodynamics is, 
of course, the partition function itself, or the free energy density,
\begin{equation}
f = - \frac{T}{V}  \ln Z(T,V)\quad .
\label{freeenergy}
\end{equation} 
All basic bulk thermodynamic observables can be derived from the 
free energy density.
In the thermodynamic limit we obtain directly the pressure, 
$p \; = \; -f$ and subsequently also other quantities like the energy 
($\epsilon$) and entropy ($s$) densities or the
velocity of sound ($c_s$), 
\begin{equation} 
{\epsilon - 3 p \over T^4} = T {{\rm d}\over {\rm d} T} 
\left(\frac{p}{T^4} \right)
\qquad , \qquad {s \over T^3} = {\epsilon + p \over T^4} 
\qquad , \qquad c_s^2 = {{\rm d} p \over {\rm d} \epsilon}\quad .
\label{eands}
\end{equation}
In the limit of infinite temperature asymptotic freedom suggest that
these observables approach the ideal gas limit for
a gas of free quarks and gluons, $\epsilon = 3p =-3f$ with $p/T^4$
given by Eq.~\ref{pSB}. Deviations from this ideal gas values have been
studied in high temperature perturbation theory. However, it was
well-known that this expansion is no longer calculable perturbatively
at ${\cal O} (g^6)$ \cite{Linde}. By now all calculable orders up to
${\cal O} (g^5\ln g)$ have been calculated \cite{Zhai}. Unfortunately
it turned out that the information gained from this expansion is 
rather limited.
The expansion shows bad convergence behaviour and suggests that it
is of use only at temperatures several orders of magnitude larger 
than the QCD transition temperature. In analytic approaches one thus
has to go beyond perturbation theory which currently is being
attempted by either using hard thermal loop resummation techniques
\cite{Bla99a,Blaizot} or perturbative dimensional reduction combined
with numerical simulations of the resulting effective 3-dimensional
theory \cite{reduced}.

In order to make use of the basic thermodynamic relations,
Eqs.~\ref{freeenergy} and \ref{eands}, in
numerical calculations on the lattice we have to go through
an additional intermediate step. The free energy density itself is not 
directly accessible in Monte Carlo calculations; e.g. only expectation 
values can be calculated easily. One thus proceeds by calculating 
differences of the free energy density at two different temperatures. 
These are obtained by taking a suitable derivative of $\ln Z$ followed 
by an integration, e.g.
\begin{equation}
\frac{f}{T^4} {\biggl |}_{T_o}^{T} \; = \; - {1\over V} \int_{T_o}^{T} 
{\rm d}x \; {\partial x^{-3} \ln Z(x,V) \over \partial x } \quad .
\label{freediv}
\end{equation}
This ansatz readily translates to the lattice. Taking derivatives
with respect to the gauge coupling, $\beta =6/g^2$, rather than the
temperature as was done in Eq.~\ref{freediv}, we obtain expectation
values of the Euclidean action which can be integrated again to give
the free energy density, 
\begin{equation}
\frac{f}{T^4} {\biggl |}_{\beta_o}^{\beta} \; = \;  N_\tau^4 
\int_{\beta_o}^{\beta}
{\rm d}\beta ' \bigl(\langle \tilde{S} \rangle - 
\langle \tilde{S} \rangle_{T=0} \bigr) \quad .
\label{freedivlat}
\end{equation} 
Here 
\begin{equation}
\langle \tilde{S} \rangle = - {1 \over N_\sigma^3 N_\tau} 
{\partial \ln Z \over \partial \beta} 
\quad ,
\label{Stilde}
\end{equation}
is calculated on a lattice of size $N_\sigma^3\times N_\tau$
and $\langle ...\rangle_{T=0}$ denotes expectation values calculated
on zero temperature lattices, which usually are approximated by symmetric
lattices with $N_\tau \equiv N_\sigma$. The lower integration limit
is chosen at low temperatures so that $f/T_o^4$ is small and may
be ignored\footnote{In the gluonic sector the relevant degrees of freedom
at low temperature are glueballs. Even the lightest ones calculated on
the lattice have large masses, $m_G \simeq 1.5$~GeV.
The free energy density thus is exponentially suppressed already close
to $T_c$. In QCD with light quarks the dominant contribution to the
free energy density comes from pions. As long as we are dealing with
massive quarks also this contribution gets suppressed exponentially.
However, in the massless limit clearly some care has to be taken with 
the normalization of the free energy density.}. 

A little bit more involved is the calculation of the energy density
as we have to take derivatives with respect to the temperature,
$T=1/N_\tau a$. On lattices with fixed temporal extent $N_\tau$ we 
rewrite this in terms of a derivative with respect to the lattice
spacing $a$ which in turn is controlled through the bare couplings
of the QCD Lagrangian, $a\equiv a(\beta, m_q)$. We thus find for
the case of $n_f$ degenerate quark  flavours of mass $m_q$
\begin{eqnarray}
{(\epsilon - 3p) \over T^4}~=~& &
N_\tau^4 \biggl[
\biggl({{\rm d}\beta(a) \over {\rm d}\ln a}\biggr) \biggl(
\langle \tilde{S} \rangle - \langle \tilde{S} \rangle_{T=0} \biggr)
\cr
&-& \biggl({{\rm d}m_q(a) \over {\rm d}\ln a}\biggr) \biggl(
\langle \bar{\chi}\chi \rangle - \langle \bar{\chi}\chi \rangle_{T=0}
\biggr)\biggr] \quad .
\label{eminus3p}
\end{eqnarray} 
An evaluation of the energy density thus e.g. requires the knowledge of
two $\beta$-functions. These may be determined by calculating two
physical observables in lattice units for given values of $\beta$
and $m_q$; for instance, the string tension,
$\sigma a^2$ and a ratio of hadron masses, $m_{PS}/m_V \equiv
m_\pi/m_\rho$.
These quantities will have to be calculated at zero temperature which 
then also allows to determine a temperature scale in physical units as
given in Eq.~\ref{tscale}. 
This forms the basis for a calculation
of the pressure as shown already in Fig.~\ref{fig:pressure}. 

The numerical calculation of thermodynamic quantities is done on finite
lattices with spatial extent $N_\sigma$ and temporal extent $N_\tau$.
In order to perform calculations close to the thermodynamic limit we want
to use a large spatial extent of the lattice. In general it has 
been found that lattices with $N_\sigma \gsim 4 N_\tau$ provide
a good approximation to the infinite volume limit. In addition,
we want to get close to the continuum limit in order to eliminate
discretization errors. Taking the continuum limit at fixed temperature
requires to perform the limit $N_\tau \rightarrow\infty$.
In order to perform this limit in a controlled way
we have to analyze in how far lattice calculations of bulk
thermodynamic observables are influenced by the introduction of a 
finite lattice cut-off, {\it i.e.} we have to understand the systematic
cut-off effects introduced through the non-zero lattice spacing.
These cut-off effects are largest in the high (infinite) temperature
limit which can be analyzed analytically in weak coupling
lattice perturbation theory. We thus will discuss this limiting 
case first.

\subsection{High temperature limit of the QCD equation of state}

In the high temperature limit bulk thermodynamic observables are
expected to approach their free gas values (Stefan-Boltzmann constants). 
In this limit cut-off effects in the pressure and in turn also in the 
energy density ($\epsilon_{\rm SB} = 3 \; p_{\rm SB}$) become 
most significant. Momenta of the order of the temperature, {\it i.e.}
short distance properties, dominate the ideal gas behaviour.

\begin{figure}[t]
\begin{center}
\hspace*{-0.4cm}
\epsfig{file=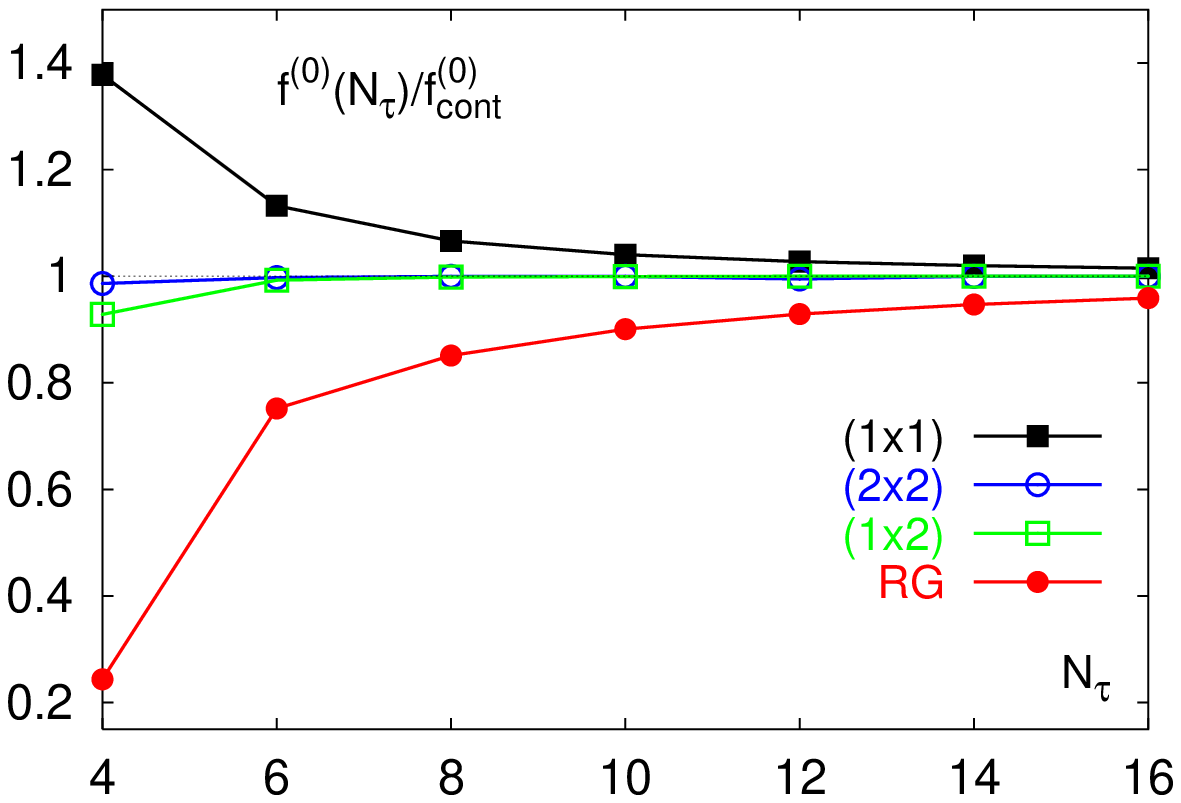,
width=64mm}\hspace*{-0.3cm}
\epsfig{file=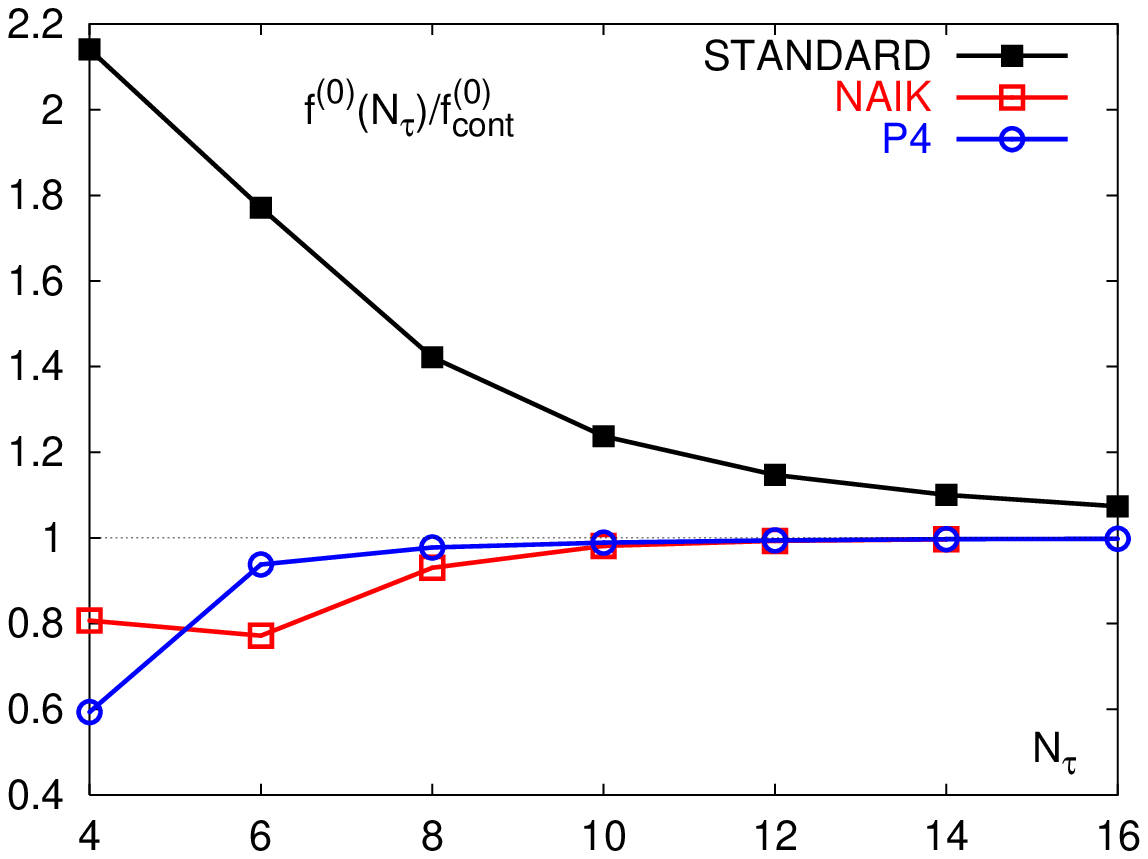,
width=62mm}
\end{center}
\caption{Cut-off dependence of the ideal gas pressure for the $SU(3)$
gauge theory (left) and several staggered fermion actions (right).
These actions are defined in the Appendix.
Cut-off effects for the Wilson fermion action are compatible with
those of the standard staggered fermion action.}
\label{fig:improvedideal}
\end{figure}
 
As discussed in Section 2 the most straightforward
lattice representation of the QCD partition function in 
terms of the standard Wilson gauge and fermion actions as well as the 
staggered fermion action leads to a systematic ${\cal O}(a^2)$ cut-off 
dependence of physical observables. At finite temperature the temperature 
itself sets the scale for these ${\cal O}(a^2)$ effects, which thus give 
rise to ${\cal O}((aT)^2 \equiv 1/N_\tau^2)$ deviations of e.g. the 
pressure from the continuum Stefan-Boltzmann value, 
\begin{equation}
{p \over T^4} {\biggl |}_{N\tau} = {p \over T^4} {\biggl |}_{\infty} 
+ {c \over N_\tau^2} + {\cal O}(N_\tau^{-4}) \quad .
\label{cutoffp}
\end{equation}
One can eliminate these leading order cut-off effects by using
improved actions which greatly reduces the cut-off dependence
in the ideal gas limit. In the Appendix we discuss
a specific set of improved gauge and fermion actions. In the 
gauge sector one may in addition to the
standard Wilson plaquette term (($1\times 1)$-action) also include 
planar 6-link 
terms in the action. This is done in the {\cal O}($a^2$) tree level
improved ($1\times 2$)-action, which eliminates the leading order 
cut-off dependence completely.
On lattices with temporal extent $N_\tau$ 
one finds for the deviation of the gluonic part of 
the pressure
\cite{Bei96},
\begin{equation}
{p_{\rm G} (N_\tau) \over p_{\rm G, SB} } = \cases{1+
{8  \over 21}  \bigl({ \pi \over N_\tau}\bigr)^2 +
{5 \over 21} \bigl({ \pi \over N_\tau}\bigr)^4 +
{\cal O} \bigl( N_\tau^{-6} \bigr) &,~~ ($1\times 1)$-action \nonumber \cr
1+ {\cal O} \bigl( N_\tau^{-4} \bigr) &,~~  ($1\times 2$)-action 
}
\label{wilsongas}
\end{equation}
A similar reduction of cut-off effects can be achieved in the fermion
sector through the use of improved actions. So far, however, improved
fermion actions, which reduce or eliminate the leading order cut-off
effects have only been constructed in the staggered fermion formulation.
The Naik action \cite{Naik}, which in addition to the ordinary one-link
term in the staggered action also includes straight three-link terms,
completely eliminates the ${\cal O}(N_\tau^{-2})$ errors on the tree
level (ideal gas limit). The p4-action discussed in the Appendix does
not eliminate this correction completely. It, however, reduces its
contribution drastically over those present in the standard staggered
action\footnote{We quote here only an approximation to the 
$N_\tau$-dependence obtained from a fit in the interval 
$10\le N_\tau \le 16$.},
\begin{equation}
{p_{\rm F} (N_\tau) \over p_{\rm F, SB} } = \cases{1+
1.57  \bigl({ \pi \over N_\tau}\bigr)^2 +
8.47 \bigl({ \pi \over N_\tau}\bigr)^4 +
{\cal O} \bigl( N_\tau^{-6} \bigr) &,~~ 1-link standard \nonumber \cr
& staggered action \cr
1+ 0.007 \bigl({ \pi \over N_\tau}\bigr)^2 +
1.07 \bigl({ \pi \over N_\tau}\bigr)^4 +
{\cal O} \bigl( N_\tau^{-6} \bigr) &,~~ p4-action } 
\label{fermiongas}
\end{equation}
as can be seen in Fig.~\ref{fig:improvedideal}. Moreover, the p4-action
has the advantage that it improves the rotational symmetry of the
fermion propagator which in turn also reduces violations of
rotational symmetry in the heavy quark potential.

Using the tree level improved gauge action in combination with the
improved staggered fermion action in numerical simulations 
at finite temperature it is possible to perform calculations with small
systematic cut-off errors already on lattices with small temporal 
extent, e.g. $N_\tau =4$ or 6. In actual calculations performed with
various actions in the pure gauge sector one finds that for 
temperatures $T\lsim 5T_c$ the cut-off dependence of thermodynamic
shows the pattern predicted by the infinite temperature perturbative
calculation. The absolute magnitude of the cut-off effects, however,
is smaller by about a factor of two. This, of course, is reassuring 
for the numerical calculations performed with light quarks, where
such a detailed systematic study of the cut-off dependence at 
present does not exist. 

\subsection{Thermodynamics of the SU(3) gauge theory}

\begin{figure}[t]
\begin{center}
\epsfig{file=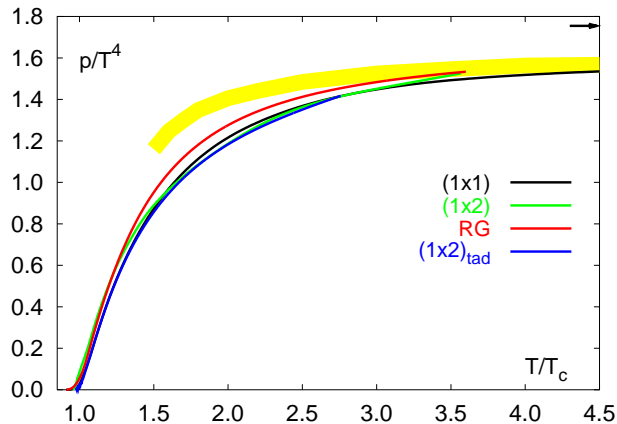,
width=85mm}
\end{center}
\caption{Pressure of the SU(3) gauge theory calculated on lattices 
with different temporal extent and extrapolated to the continuum limit.
Shown are results from calculations with the standard Wilson 
$(1\times 1)$-action \cite{Boy96} and several improved actions 
\cite{Oka99,Bei99}, which are defined in the Appendix. The broad band 
shows the approximately self-consistent HTL calculation of \cite{Bla99}.}
\label{fig:su3eos}
\end{figure}

Before entering a discussion of bulk thermodynamics in two and three 
flavour QCD it is worthwhile to discuss some results on the equation 
of state in the heavy quark mass limit of QCD -- the SU(3) gauge theory. 
In this case the temperature dependence of the pressure and energy
density has been studied in great detail, calculations with the standard
action \cite{Boy96} and various improved actions \cite{Pap96,Oka99,Bei99} 
have been
performed, the cut-off dependence has explicitly been analyzed through
calculations on lattices with varying temporal extent $N_\tau$ and
results have been extrapolated to the continuum limit. In 
Fig.~\ref{fig:su3eos} we show some results for the pressure obtained from 
such detailed analyzes with different actions \cite{Boy96,Oka99,Bei99}. 
This figure shows the basic features of the temperature dependence
of bulk thermodynamic quantities in QCD, which also carry over
to the case of QCD with light quarks. The pressure stays small for
almost all temperatures below $T_c$; this is expected, as the only
degrees of freedom in the low temperature phase are glueballs which
are rather heavy and thus lead to an exponential suppression of
pressure and energy density at low temperature. Above $T_c$ the pressure
rises rapidly and reaches about 70\% of the asymptotic ideal gas
value at $T=2\; T_c$. For even larger temperatures the approach to 
this limiting value proceeds rather slowly. In fact, even at $T\simeq
4\; T_c$ deviations from the ideal gas value are larger than 10\%. 
This is too much to be understood in terms of weakly interacting
gluons as they are described by ordinary high temperature
perturbation theory \cite{Zhai}. Even at these high temperatures
non-perturbative effects have to be taken into account which may
be described in terms of interactions among quasi-particles
\cite{Bla99a,quasi}. In Fig.~\ref{fig:su3eos} we show 
the result of a self-consistent HTL resummation \cite{Bla99}, which 
leads to good agreement with the lattice calculations for $T\gsim 3T_c$.
Other approaches \cite{reduced,quasi} reach a similarly
good agreement in the high temperature regime.

\begin{figure}[t]
\begin{center}
\epsfig{bbllx=92,bblly=106,bburx=504,bbury=685,
file=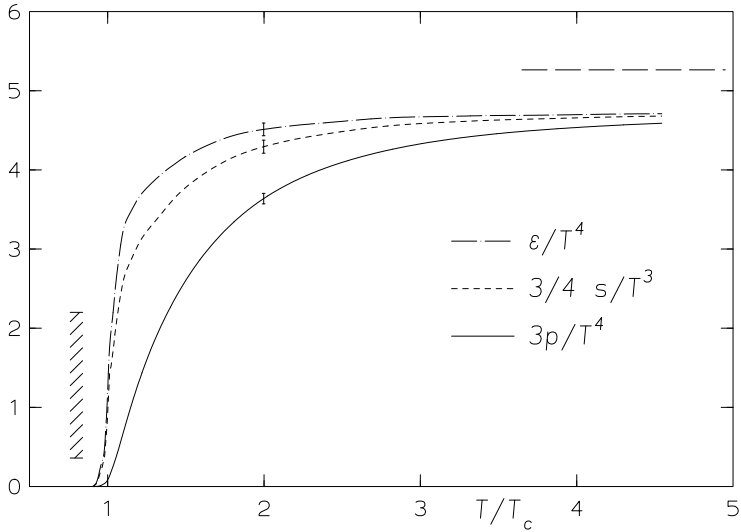, width=70mm,angle=-90}
\end{center}
\caption{Energy density, entropy density and pressure of 
the SU(3) gauge theory calculated on lattices 
with different temporal extent and extrapolated to the continuum limit.
The dashed band indicates the size of the latent heat gap in 
energy and entropy density}
\label{fig:su3etsp}
\end{figure}

Compared to the pressure the energy density rises much more
rapidly in the vicinity of $T_c$. In fact, as the transition is
first order in the $SU(3)$ gauge theory the energy density is
discontinuous at $T_c$ with a latent heat of about $1.5 T_c^4$
\cite{Bei97}. In Fig.~\ref{fig:su3etsp} we show results for 
the energy density, entropy density and the pressure obtained
from calculations with the Wilson action which have been 
extrapolated to the continuum limit \cite{Boy96}.
The delayed rise of the pressure compared to that of the energy density
has consequences for the velocity of sound in the QCD plasma in 
the vicinity of $T_c$. It is substantially smaller than in the high
temperature ideal gas limit.

\subsection{Flavour dependence of the QCD equation of state}

As shown in Eq.~\ref{freedivlat} the pressure in QCD
with light quarks can be calculated along the same line as in
the pure gauge sector. Unlike in the pure gauge case it, however,
will be difficult to perform calculations on lattices with large 
temporal extent. In fact, at present all calculations of the
equation of state are restricted to lattices with $N_\tau =4$ and 6
\cite{Kar00a,Ber97b,Ali01b}.
The use of an improved fermion action thus seems to be even more 
important in this case. Of course, an additional problem arises
from insufficient chiral properties of staggered and Wilson fermion
actions. This will mainly be of importance in the low temperature
phase and in the vicinity of the transition temperature.
The continuum extrapolation thus will be more involved in the 
case of QCD with light quarks than in the pure gauge theory 
and we will have to perform calculations closer to 
the continuum limit. Nonetheless, in particular for small number of
flavours, we may expect that the flavour symmetry breaking only has a 
small effect on the overall magnitude of bulk thermodynamic observables. 
After all, for $n_f=2$, the pressure of an ideal massless pion gas
contributes less than 10\% of that of an ideal quark-gluon gas in the 
high temperature limit. For our discussion of bulk thermodynamic observables 
the main source for lattice artifacts thus still seems to arise from the short 
distance cut-off effects, which we have to control. Additional confidence 
in the numerical results can be gained by comparing simulations 
performed with different fermion actions.

\begin{figure}[t]
\begin{center}
\epsfig{file=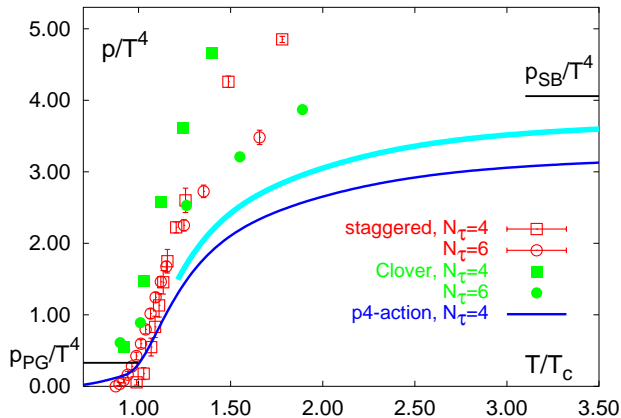,
width=85mm}
\end{center}
\caption{The pressure in two flavour QCD calculated with unimproved
gauge and staggered fermion actions (open symbols) \cite{Ber97b}, 
RG-improved gauge
and clover improved Wilson action (full symbols) \cite{Ali01b} and 
the p4-action (improved gauge and improved staggered fermions, see
Appendix) (full line) \cite{Kar00a}. 
The dashed band estimates the results in the continuum
limit as described in the text. The horizontal lines to the right and left
show the Stefan-Boltzmann values for an ideal pion gas and a free quark-gluon
gas, respectively.}
\label{fig:pressure_nf2_std_p4}
\end{figure}

The importance of an improved lattice action, which leads to
small cut-off errors at least in the high temperature ideal gas limit 
is apparent from Fig.~\ref{fig:pressure_nf2_std_p4}, where we compare
the results of a calculation of the pressure in 2-flavour QCD
performed with unimproved gauge and staggered fermion actions
\cite{Ber97b} and the RG-improved gauge and clover improved Wilson action
\cite{Ali01b} with results obtained with the p4-action discussed 
in the Appendix.  At temperatures above $T\simeq 2\; T_c$ these 
actions qualitatively reproduce the cut-off effects calculated
analytically in the infinite temperature limit (see Section 6.1). 
In particular, it is evident that also the Clover improved  
Wilson action leads to an overshooting of the continuum ideal
gas limit. This is expected as the Clover term in the Wilson action
does eliminate ${\cal O}(ag^2)$ cut-off effects but does not improve the 
high temperature ideal gas limit, which is ${\cal O}(g^0)$. The clover
improved Wilson action thus leads to the same large ${\cal O}(a^2)$ cut-off 
effects as the unimproved Wilson action. 
The influence of cut-off effects in bulk thermodynamic observables
thus is similar in calculations with light quarks and in the 
$SU(3)$ gauge theory. This observation may also help to estimate the
cut-off effects still present in current calculations with light
quarks. In particular, we know from the analysis performed in the pure 
gauge sector that in the interesting temperature regime of a few times 
$T_c$ the cut-off dependence seems to be about a factor two smaller than
calculated analytically in the infinite temperature limit; we may 
expect that this carries over to the case of QCD with light quarks. 
This is the basis for the estimated 
continuum extrapolation of the $n_f=2$ results shown as a dashed band
in Fig.~\ref{fig:pressure_nf2_std_p4}.  

In Fig.~\ref{fig:pressure} we have already shown result for the pressure
calculated in QCD with different number of flavours. This figure clearly 
shows that the transition region shifts to smaller temperatures as the 
number of degrees of freedom is increased.
Such a conclusion, of course, requires the determination of a
temperature scale that is common to all {\it QCD-like} theories which
have a particle content different from that realized in nature. We have
determined this temperature scale by assuming that the string tension is 
flavour and quark mass independent. This assumption is supported
by the observation that already in the heavy quark mass limit the
string tension calculated in units of quenched hadron masses, 
e.g. $m_\rho /\sqrt{\sigma} = 1.81~(4)$ \cite{Wit97}, is in good
agreement with values required in QCD phenomenology, $\sqrt{\sigma}
\simeq 425~{\rm MeV}$.

\begin{figure}
\begin{center}
\epsfig{file=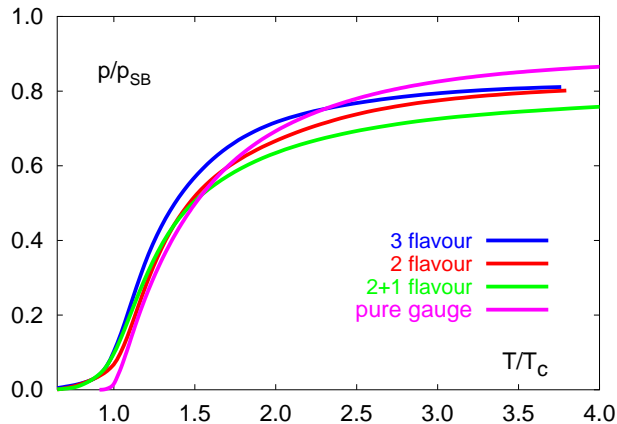,
width=85mm}
\end{center}
\caption{The pressure in units of the ideal gas pressure for 
the $SU(3)$ gauge theory and QCD with various number of flavours.
The latter calculations have been performed on lattices with temporal
extent $N_\tau =4$ using the p4-action defined in the Appendix. Results
are not yet extrapolated to the continuum limit.}
\label{fig:p_psb}
\end{figure}

At high temperature the magnitude of $p/T^4$ clearly
reflects the change in the number of light degrees of freedom present
in the ideal gas limit. When we rescale the pressure by the corresponding 
ideal gas values it becomes, however, apparent that the overall pattern 
of the temperature dependence of $p/T^4$ is quite similar in all cases.
This is shown in Fig.~\ref{fig:p_psb}. In particular, when one takes
into account that a proper continuum extrapolation in QCD with light
quarks is still missing this agreement achieved with improved staggered
fermions is quite remarkable. 

\begin{figure*}[t]
\begin{center}
\hspace*{0.3cm}\epsfig{
file=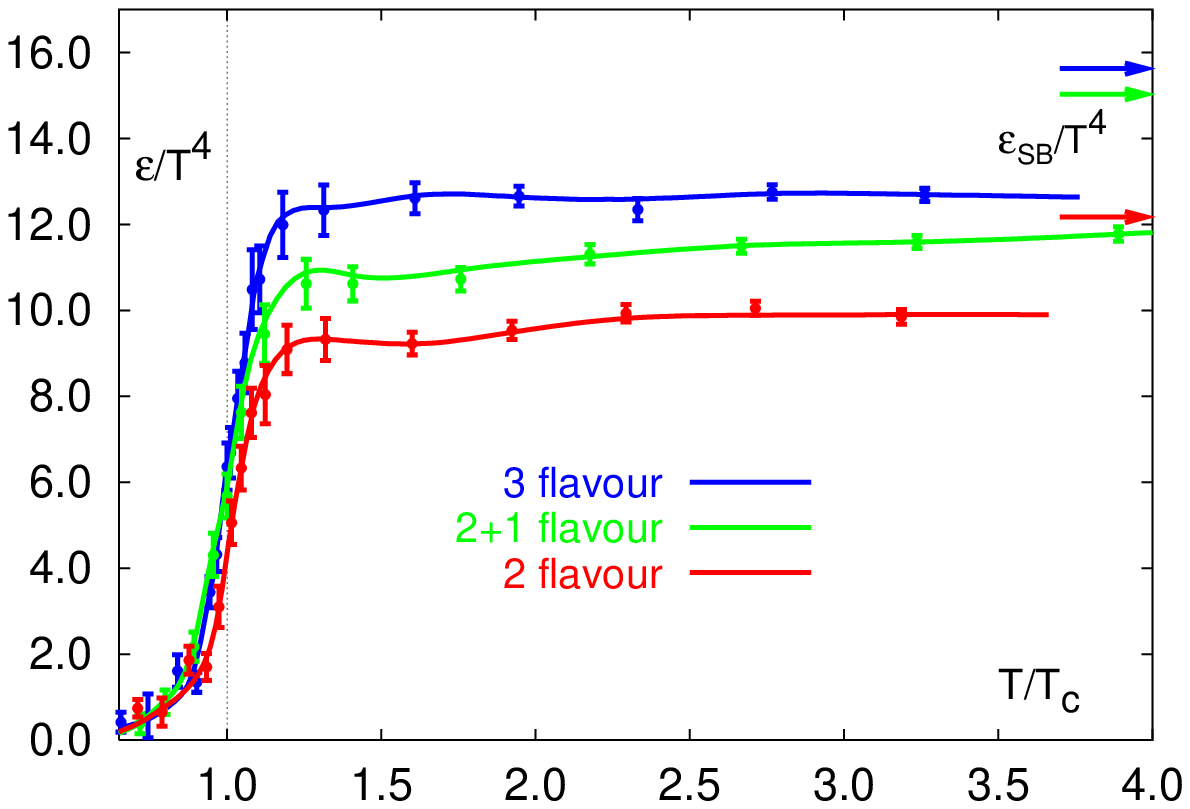,width=82mm}

\epsfig{bbllx=25,bblly=88,bburx=531,bbury=445,
file=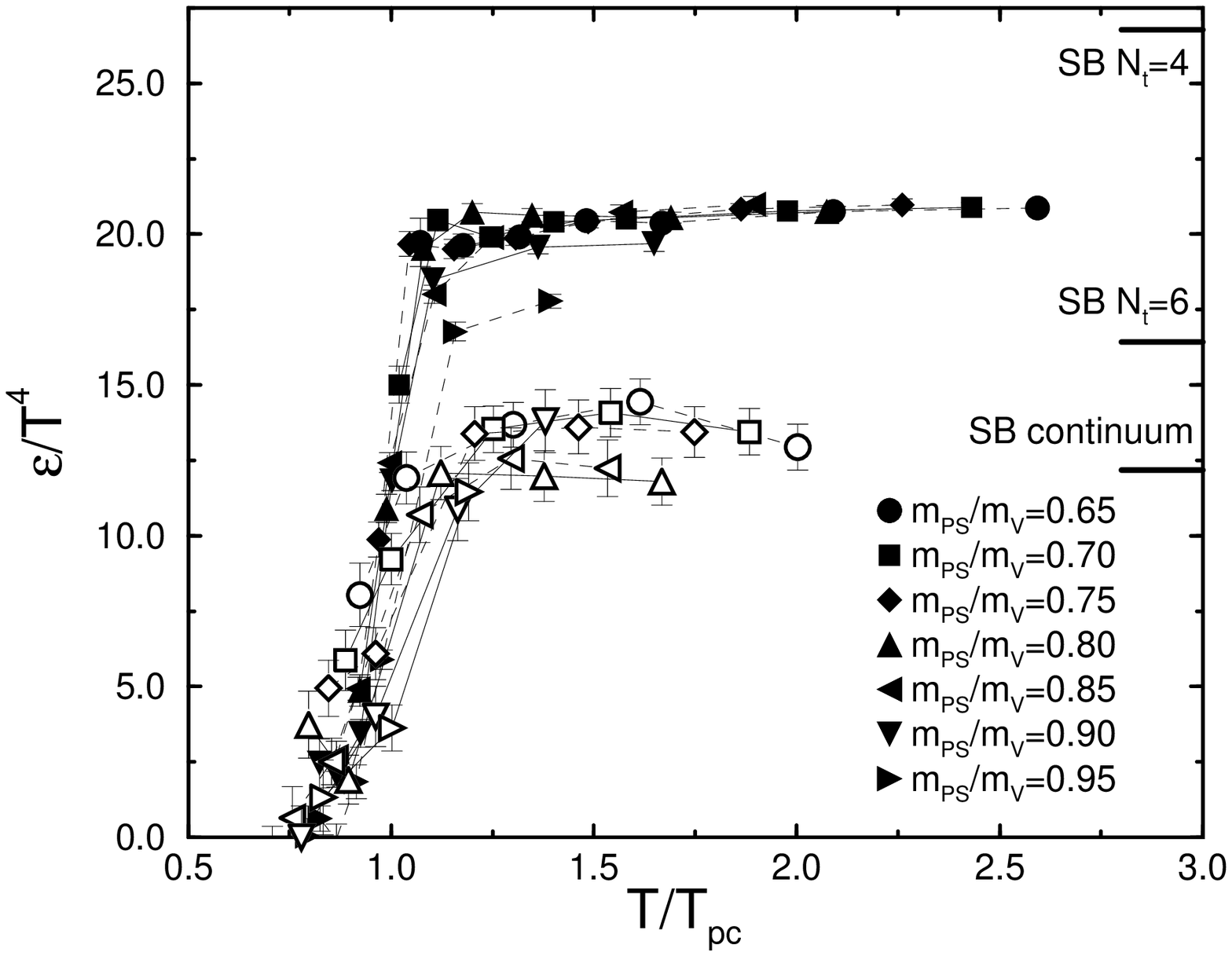,width=81mm}
\end{center}
\vskip 0.8truecm
\caption{The energy density in QCD. The upper (lower) figure shows
results
from a calculation with improved staggered \cite{Kar00a} (Wilson
\cite{Ali01b}) fermions on lattices
with temporal extent $N_\tau=4$ ($N_\tau=4,~6$). 
The staggered fermion calculations have been performed for a
pseudo-scalar to vector meson mass ratio of $m_{PS}/m_V =0.7$.}
\label{fig:energy_nf}
\end{figure*}

We also note that the pressure at low
temperature is enhanced in QCD with light quarks compared to the pure
gauge case. This is an indication for the contribution of hadronic
states, which are significantly lighter than the heavy glueballs of 
the $SU(3)$ gauge theory. 
This behaviour is even more clearly visible in the behaviour of the energy 
density which is shown in Fig.~\ref{fig:energy_nf}, where we show results
obtained with improved staggered\footnote{The
figure for staggered fermions is based on data from 
Ref.~\cite{Kar00a}. Here a contribution to $\epsilon/T^4$ which
is proportional to the bare quark mass and vanishes in the chiral 
limit is not taken into account.} and Wilson \cite{Ali01b} fermions.
We note that these calculations yield consistent estimates for the
energy density at $T_c$, 
\begin{equation}
\epsilon_c \simeq (6 \pm 2) T_c^4 \quad .
\label{ec}
\end{equation}

This estimate for $\epsilon_c/T_c^4$, which also is consistent with results 
obtained  from calculations with a standard staggered fermion action
\cite{Ber97b}, is an order of magnitude larger than the critical 
value on the hadronic side of the transition in the pure gauge theory
(see Fig.~\ref{fig:su3etsp}). It is, however, interesting to note that 
when we convert this result for $\epsilon_c$ in physical units, 
[MeV/fm$^3$], this difference gets to a large extent compensated by the 
shift in $T_c$ to smaller values. 
When going from the infinite quark mass limit to
the light quark mass regime the QCD transition thus seems to take
place at compatible values of the energy density, $\epsilon_c \simeq
(0.5 - 1){\rm GeV/fm}^3$. The largest uncertainty on this number at 
present arises from uncertainties on the value of $T_c$ (see next
section). However, also the magnitude of $\epsilon_c/T_c^4$ still
has to be determined more accurately. Here two competing effects
will be relevant. On the one hand we expect $\epsilon_c/T_c^4$ to 
increase with decreasing quark masses, {\it i.e.} closer to the chiral
limit. On the other hand, it is likely
that finite volume effects are similar to those in the pure gauge 
sector, which suggests that $\epsilon_c/T_c^4$ will still decrease 
closer to the thermodynamic limit, {\it i.e.} for $N_\sigma \rightarrow
\infty$. 

In the 2-flavour calculations performed with improved Wilson fermions
\cite{Ali01b}  the pressure and energy density have been calculated 
for several values of the quark mass, which corresponds to different
ratios of the pseudo-scalar to vector meson mass $m_{PS}/m_V$. The
results show no significant quark mass dependence up to $m_{PS}/m_V
\simeq 0.9$. This meson mass ratio corresponds to pseudo-scalar
meson masses of about 1.5~GeV, which is somewhat larger than the
$\Phi$-meson mass. This suggests that the corresponding  
quark mass is compatible with that of the strange quark. 
The approximate quark mass independence of the equation of state 
observed in the high temperature phase thus is consistent with
our expectation that quark mass effects should become significant
only when the quark masses get larger than the temperature.

\section{The Critical Temperature of the QCD Transition}

As discussed in Section 3 the transition to the high temperature 
phase is continuous and non-singular for a large range of quark 
masses. Nonetheless, for all quark masses this transition proceeds 
rather rapidly in a small temperature interval. A definite 
transition point thus can be identified, for instance through the
location of peaks in the susceptibilities of the Polyakov loop
or the chiral condensate defined in Eq.~\ref{sus}. For a given
value of the quark mass one thus determines pseudo-critical 
couplings, $\beta_{pc}(m_q)$, on a lattice with temporal extent $N_\tau$.
An additional calculation of an experimentally or phenomenologically
known observable at zero temperature, e.g. a hadron mass or the string 
tension, is still needed to determine the transition temperature
from Eq.~\ref{tscale}. In the pure gauge theory the transition temperature,
again has been analyzed in great detail and the influence of cut-off
effects has been examined through calculations on different size 
lattices and with different actions. From this one finds for the 
critical temperature of the first order phase transition in the pure 
$SU(3)$ gauge theory,
\begin{eqnarray}
\underline{\rm SU(3)~gauge~theory:} && T_c/\sqrt{\sigma} = 0.637\pm
0.005 \nonumber \\
&& T_c = (271\pm 2) \; {\rm MeV}
\label{tcsu3}
\end{eqnarray} 

\begin{figure}[t]
\begin{center}
\epsfig{file=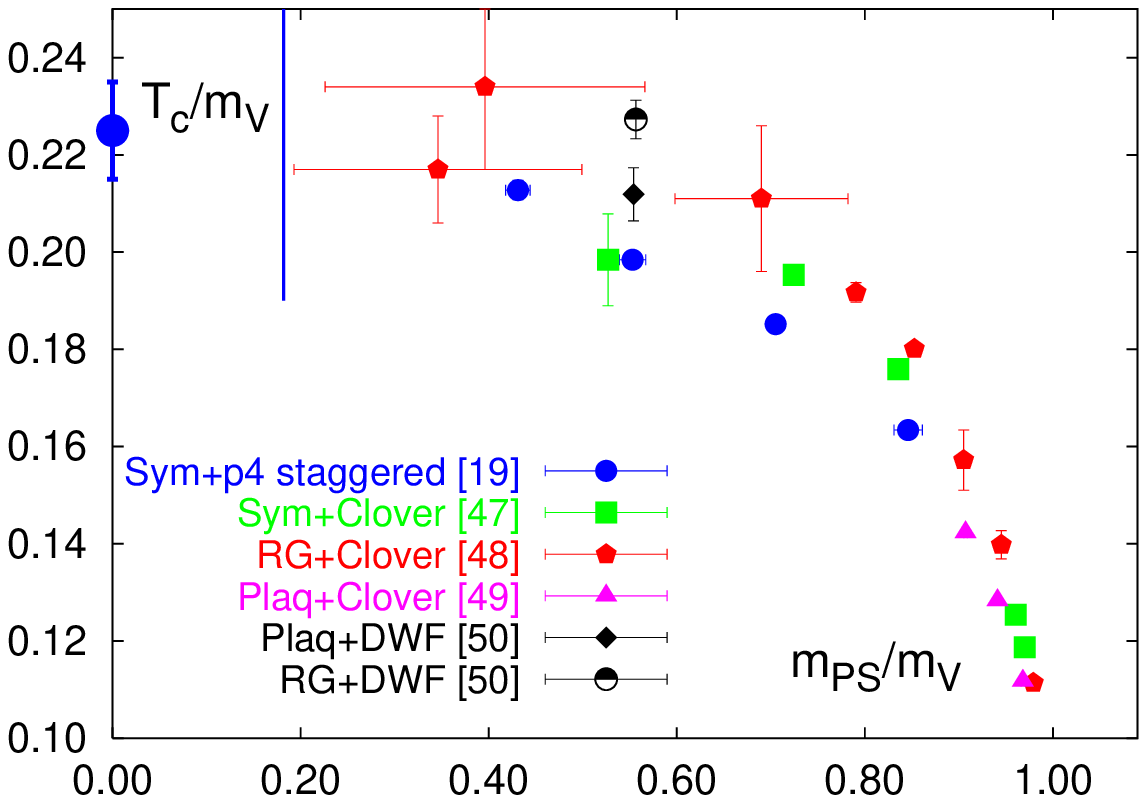,width=85mm}

\epsfig{file=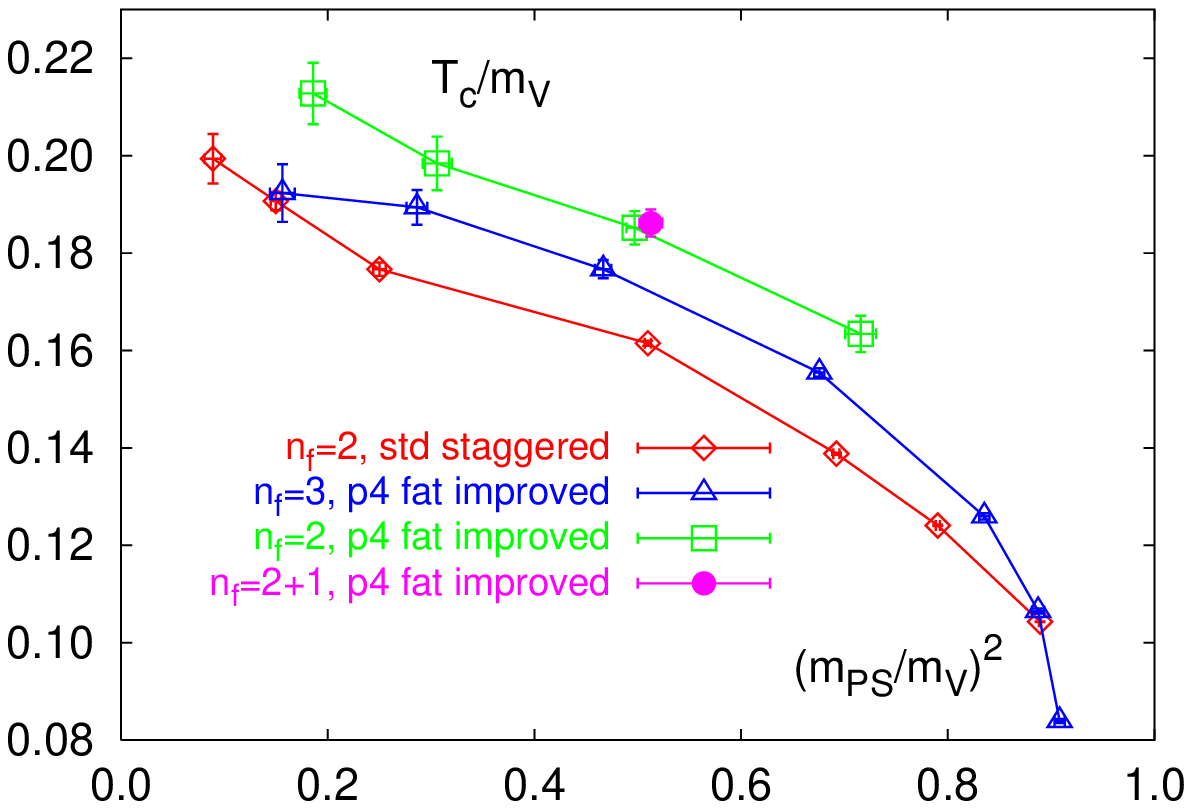,width=82mm}
\end{center}
\caption{Transition temperatures in units of $m_V$. The upper figure
shows 
a collection of results obtained for 2-flavour QCD with various fermion
actions while in the lower figure we compare results obtained in
2 and 3-flavour QCD with the p4-action described in the Appendix. 
All results are from simulations on lattices with temporal
extent $N_\tau = 4$. The large dot drawn for $m_{PS}/m_V =0$
indicates the result of chiral extrapolations based on
calculations with improved Wilson \cite{Ali01} as well as improved
staggered \cite{Kar01} fermions.
The vertical line in the upper figure shows the location of the physical 
limit, $m_{PS}\equiv m_\pi = 140~{\rm MeV}$.
}
\vskip -0.2truecm
\label{fig:tcnf2}
\end{figure}

Already the early calculations for the transition temperature with light
quarks \cite{Bit91,Ber97} indicated that the inclusion of light quarks
leads to a significant decrease of the transition temperature. However,
these early calculations, which have been performed with standard 
Wilson \cite{Bit91} and staggered \cite{Ber97} fermion actions, also 
led to significant discrepancies in the results for $T_c$ as well
as the order of the transition. These differences strongly diminished
in the newer calculations which are based on improved Wilson fermions
(Clover action) \cite{Ber97,Ali01,Edw99}, domain wall fermions \cite{norman}
as well as improved staggered fermions (p4-action) \cite{Kar01}. A
compilation of these newer results is shown in Fig.~\ref{fig:tcnf2} for
various values of the quark masses. In order to compare calculations performed
with different actions the results are presented in terms of a {\it physical
observable}, the meson mass ratio $m_{PS}/m_V$. In Fig.~\ref{fig:tcnf2}a we
show $T_c/m_V$ obtained for 2-flavour QCD while Fig.~\ref{fig:tcnf2}b
gives a comparison of results obtained with improved
staggered fermions \cite{Kar01} for 2 and 3-flavour QCD.
Also shown there is a result for the case of (2+1)-flavour QCD, {\it i.e.}
for two light and one heavier quark flavour degree of freedom.
Unfortunately the quark masses in this latter case are still
too large to be compared directly with the situation realized in nature.
We note however, that the results obtained so far suggest that the transition
temperature in (2+1)-flavour QCD is close to
that of 2-flavour QCD. The 3-flavour theory, on the other hand, leads to
consistently smaller values of the critical temperature,
$T_c(n_f=2)-T_c(n_f=3) \simeq 20$~MeV. The extrapolation of the transition
temperatures to the chiral limit gave
\begin{eqnarray}
\underline{\rm 2-flavour~ QCD:} &&
T_c  = \cases{(171\pm 4)\; {\rm MeV}, &  clover-improved
Wilson \nonumber \cr
~& fermions \cite{Ali01} \cr
(173\pm 8)\; {\rm MeV}, & improved staggered  \nonumber \cr
~& fermions  \cite{Kar01}}
\nonumber \\
\underline{\rm 3-flavour~ QCD:} &&
T_c  = \; \; \;\;
(154\pm 8)\; {\rm MeV}, \; \; \hspace*{0.1cm}  
{\rm improved~ staggered} \nonumber \\
~&& \hspace*{3.6cm}{\rm fermions}~[19]  
\nonumber
\end{eqnarray}
Here $m_\rho$ has been used to set the scale for $T_c$.  
Although the agreement between results obtained with Wilson and
staggered fermions is striking, one should bear in mind that all
these results have been obtained on lattice with temporal extent $N_\tau =4$,
{\it i.e.} at rather large lattice spacing, $a\simeq 0.3$~fm.
Moreover, there are uncertainties involved in the ansatz used to
extrapolate to the chiral limit. We thus estimate that the systematic
error on the value of $T_c /m_\rho$ still is of similar magnitude as
the purely statistical error quoted above.
 
We note from Fig.~\ref{fig:tcnf2} that $T_c/m_V$ drops with increasing
ratio $m_{PS}/m_V$, {\it i.e.} with increasing quark mass. This may
not be too surprising as $m_V$, of course, does not take on 
the physical $\rho$-meson mass value as long as $m_{PS}/m_V$ did
not reach is physical value (vertical line in Fig.~\ref{fig:tcnf2}a).
In fact, we know that $T_c/m_V$ will approach zero for $m_{PS}/m_V =1$
as $T_c$ will stay finite and take on the value calculated in the
pure $SU(3)$ gauge theory whereas $m_V$ will diverge in the heavy
quark mass limit. Fig.~\ref{fig:tcnf2} thus does not yet allow to
quantify how $T_c$ depends on the quark mass.
A simple percolation picture for the QCD transition
would suggest that $T_c (m_q)$ or better $T_c(m_{PS})$ will increase
with increasing $m_q$; with increasing $m_q$
also the hadron masses increase and it becomes more difficult to
excite the low lying hadronic states. It thus becomes more difficult to
create a sufficiently high particle/energy density in the
hadronic phase that can trigger a phase (percolation) transition. Such a
picture also follows from chiral model calculations \cite{chiralmodel}.

As argued previously we should express $T_c$ in units of an observable, 
which itself is not dependent on $m_q$; the string tension (or also a
quenched hadron mass) seems to be suitable for this purpose. 
In fact, this
is what tacitly has been assumed when one converts the critical
temperature of the SU(3) gauge theory $T_c/\sqrt{\sigma} \simeq 0.63$ into
physical units as has also been done in Eq.~\ref{tcsu3}.
\begin{figure}[t]
\begin{center}
\epsfig{file=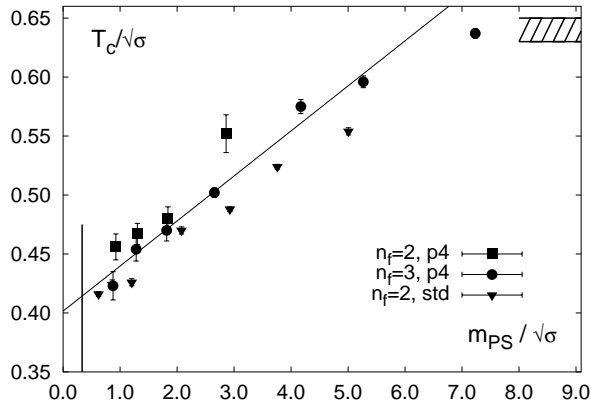,width=82mm}
\end{center}
\caption{The transition temperature in 2 (filled squares) and 3 (circles)
flavour  QCD versus $m_{PS}/\sqrt{\sigma}$ using an improved staggered
fermion action (p4-action). Also shown are results for 2-flavour QCD
obtained with the standard staggered fermion action (open squares).
The dashed band indicates the uncertainty on $T_c/\sqrt{\sigma}$ in the
quenched limit. The straight line is the fit given in Eq.~\ref{tcfit}.}
\label{fig:tc_pion}
\end{figure}

To quantify the quark mass dependence of the transition temperature
one may express $T_c$ in units of $\sqrt{\sigma}$.
This ratio is shown in Fig.~\ref{fig:tc_pion} as a function of
$m_{PS} / \sqrt{\sigma}$. As can be seen the transition
temperature starts deviating from the quenched values for $m_{PS}\;
\lsim \; (6-7)\sqrt{\sigma}\simeq 2.5~{\rm GeV}$. We also note that the
dependence of $T_c$ on $m_{PS}/\sqrt{\sigma}$
is almost linear in the entire mass interval.
Such a behaviour might, in fact, be expected
for light quarks in the vicinity of
a $2^{nd}$ order chiral transition where the dependence of the 
pseudo-critical temperature on the mass of the Goldstone-particle 
follows from the scaling relation, Eq.~\ref{pcline}, 
\begin{equation}
T_c(m_{\pi}) - T_c(0) \sim m_{\pi}^{2/\beta\delta} ~.
\end{equation}
For 2-flavour QCD the critical indices are expected to
belong to the universality class of 3-d, $O(4)$
symmetric spin models and one thus would indeed expect $1/\beta\delta=0.55$.
However, this clearly cannot be the origin of the quasi linear
behaviour which is observed for rather large hadron masses and seems
to be independent of $n_f$.  
Moreover, unlike in chiral models \cite{chiralmodel} the dependence
of $T_c$ on $m_{PS}$ turns out to be rather weak. The line shown in
Fig.~\ref{fig:tc_pion} is a fit to the 3-flavour data, which gave
\begin{equation}
\label{tcfit}
\biggl({T_c \over \sqrt{\sigma}} \biggr)_{m_{PS}/\sqrt{\sigma}} = 
\biggl({T_c \over \sqrt{\sigma}} \biggr)_0 +
0.04(1)\; \biggl({m_{PS} \over \sqrt{\sigma}} \biggr) \quad .
\end{equation}

It thus seems that the transition temperature does not react
strongly on changes of the lightest hadron masses. This favours
the interpretation that the contributions of heavy resonance masses
are equally important for the occurrence of the transition. In fact,
this also can explain why the transition still sets in at quite 
low temperatures even when all hadron masses, including the pseudo-scalars,
attain masses of the order of  1~GeV or more. Such an interpretation also
is consistent with the weak quark mass dependence of the critical
energy density we found from the analysis of the QCD equation of 
state in the previous section. 

For the quark masses currently used in lattice calculations
a resonance gas model combined with a percolation criterion
thus provides an appropriate to describe the thermodynamics
close to $T_c$. It remains to be seen whether the role of the 
light meson sector becomes more dominant when we get closer
to the chiral limit.      

\section{Finite Density QCD}

Finite density calculations in QCD are affected by the well known
sign problem, {\it i.e.} the fermion determinant appearing in the
QCD partition function, Eq.~\ref{zsf}, becomes complex for
non-zero values of the chemical potential $\mu$ and thus prohibits the
use of conventional numerical algorithms. The most detailed studies
of this problem have so far been performed using the Glasgow 
algorithm \cite{Bar98},
which is based on a fugacity expansion of the grand canonical partition
function at non-zero $\mu$,

\begin{equation}
Z_{GC}(\mu/T,T,V) = \sum_{B=-\alpha V}^{\alpha V} z^B Z_B(T,V) ~,
\label{gcp}
\end{equation}
where $z=\exp{(\mu /T)}$ is the fugacity and $Z_B$ are the canonical
partition functions for fixed quark number $B$; $\alpha = 3,6$ for
one species of staggered or Wilson fermions, respectively. 
After introducing a complex chemical potential in $Z_{GC}$
the canonical partition functions can be obtained via a Fourier
transformation\footnote{The use of this ansatz for the calculation of
canonical partition functions as expansion coefficients for $Z_{GC}$ 
has been discussed in \cite{Has92,Alf99}. A new approach has been
suggested recently, which combines simulations with imaginary chemical
potential with an analytic continuation based on the Ferrenberg-Swendsen
multi-histogram method \cite{Fod01}.},
\begin{eqnarray}
\label{canonical}
Z_B(T,V) &=& {1\over 2\pi} \int_0^{2\pi}{\rm d}\phi \; {\rm e}^{i\phi
B}\; Z_{GC} (i\phi,T,V)~ \\
&\equiv& \int \prod_{n \nu} \D U_{n,\nu} a_B \E^{-\beta S_{G}}   
\quad ,
\end{eqnarray}
with
\begin{equation}
a_B = {1\over 2\pi} \int_0^{2\pi}{\rm d}\phi \; {\rm e}^{i\phi
B}\; (\det Q^{F}(m_q, i\phi))^{n_f/4} \quad .
\label{aB}
\end{equation}
One thus may
evaluate the canonical partition functions as expectation values
with respect to a trial partition function that can be handled 
numerically, for instance the partition function of the pure $SU(3)$ gauge 
theory,
\begin{equation} 
Z_{GC}(\mu/T,T,V) = Z_{\rm SU(3)} \sum_{B=-\alpha V}^{\alpha V} z^B 
\langle a_B \rangle_{\rm SU(3)} \quad .
\end{equation} 
However, this approach so far did not overcome the severe numerical 
difficulties. Like other approaches it suffers from the problem that
expectation values have to be calculated with respect to another ensemble
so that the importance sampling which is at the heart of every numerical
approach samples the wrong region of phase space and thus may become
quite inefficient.

It thus may be helpful to approach the finite density problems
from another perspective. A reformulation of the original ansatz
may lead to a representation of the partition function which, in
the ideal case, would require the averaging over configurations
with strictly positive weights only, or at least would lead to a
strong reduction of configurations with negative weights.

An alternative formulation of finite density QCD is given in terms
of canonical rather than grand canonical partition functions
\cite{Mil87},
{\it i.e.} rather than introducing a non-zero chemical potential through
which the number density is controlled one introduces directly a
non-zero baryon number (or quark number $B$) through Eq.~\ref{canonical} 
from which the baryon number density on lattices of size 
$N_\sigma^3 \times N_\tau$ is obtained as
${n_B / T^3} = {B\over  3} ({N_\tau / N_\sigma})^3$.
Also this formulation is by no means easy to use in general, {\it i.e.}
for QCD with light quarks. In particular, it also still suffers
from a sign problem. It, however, leads to a quite natural and
useful formulation of the quenched limit of QCD at non-zero
density \cite{Eng99b} which may be a good starting point for generalizing
this approach to finite values of the quark mass. In the following
we briefly outline the basic ideas of this approach.

%
\subsection{Quenched limit of finite density QCD}

\begin{figure*}[t]
\hspace*{-0.2cm}\epsfig{file=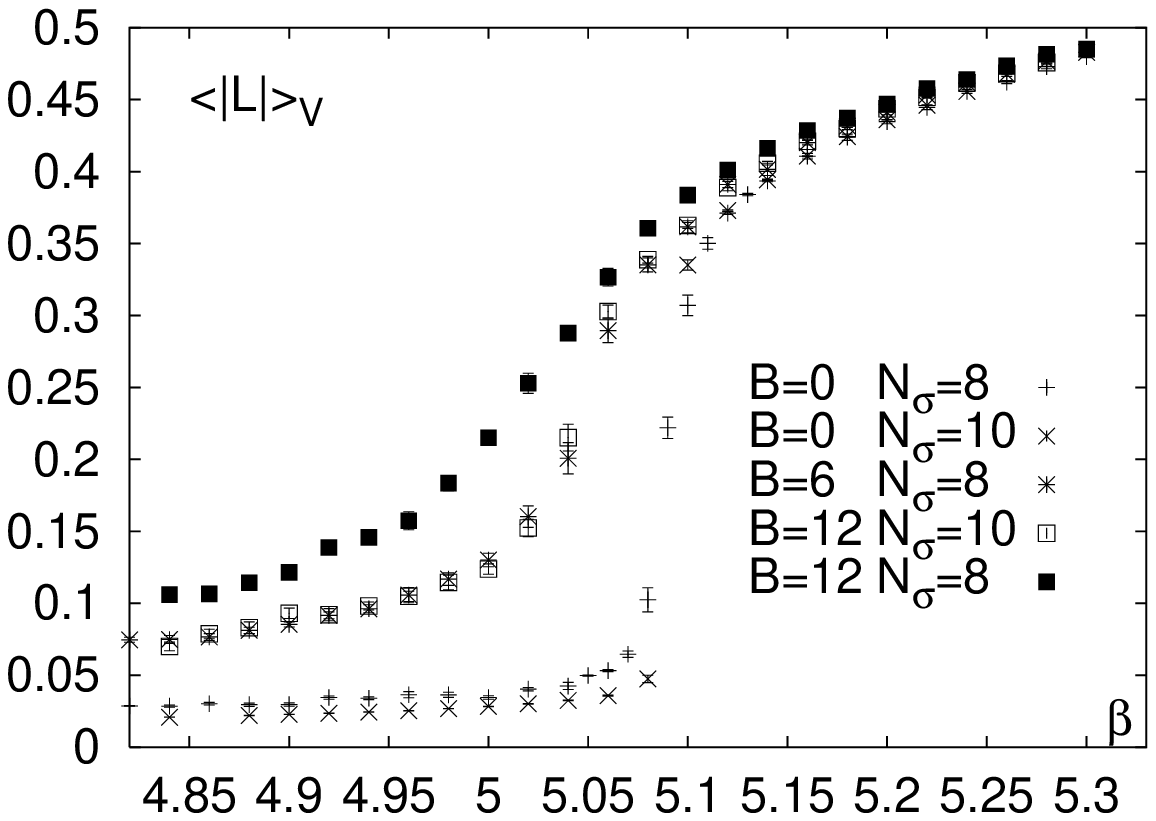,width=62mm}
\hspace*{-0.3cm}\epsfig{file=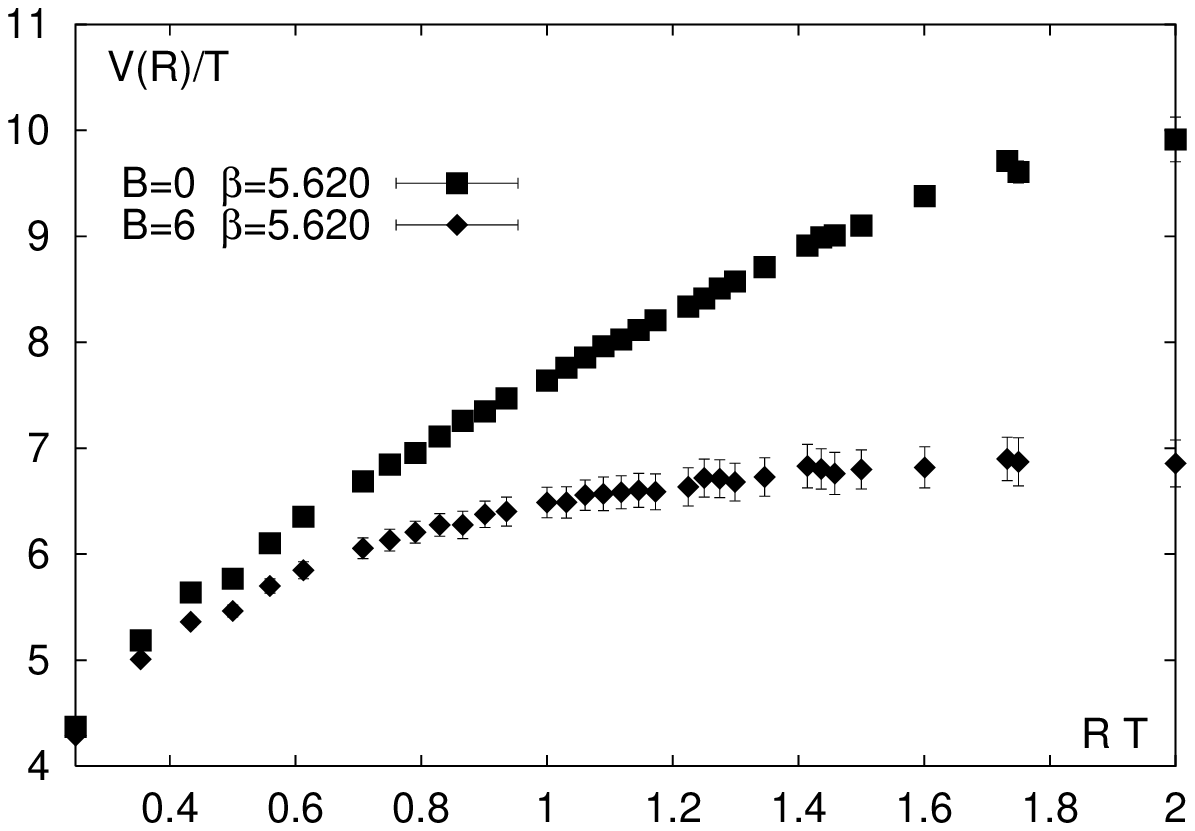,width=62mm}
\caption{Polyakov loop expectation value (left) calculated on
$N_\sigma^3 \times 2$ and the heavy quark
potential (right) calculated on $16^3\times 4$ lattices in quenched QCD
at zero and non-zero baryon number, $B/3$.}
\label{fig:density}
\end{figure*}

It had been noticed early that the
straightforward replacement of the fermion determinant by a constant
does not lead to a meaningful static limit of QCD \cite{Bar86}. In
fact, this simple replacement corresponds to the static limit
fermion flavours carrying baryon number $B$
and $-B$, respectively \cite{Ste96}. This should not be
too surprising. When one starts with QCD at a non-zero baryon number
and takes the limit of infinitely heavy quarks something should be
left over from the determinant that represents the objects that carry
the baryon number. In the canonical formulation this becomes obvious.
For $m_q\rightarrow \infty$ one ends up with a partition function, which
for baryon number $B/3$ still includes the sum over products of $B$
Polyakov loops, {\it i.e.} the static quark propagators which carry
the baryon number \cite{Eng99b}. This limit also has some
analogy in the grand canonical formulation where the coupled limit
$m_q,\mu\rightarrow \infty$ with $\exp{(\mu)}/2m_q$
kept fixed has been performed \cite{Sta91,Blu96}\footnote{This is a
well known limit in statistical physics. When deriving the
non-relativistic gas limit from a relativistic gas of particles
with mass $\bar{m}$, the rest mass is splitted off from the chemical
potential, $\mu \equiv \mu_{nr} + \bar{m}$,
in order to cancel the corresponding rest mass term in the particle
energies. On the lattice $\bar{m} = \ln (2m_q)$ for large bare quark
masses.}.

As the baryon number is carried by the rather heavy nucleons
in the confined phase of QCD we may expect that it is quit
reasonable to approximate them by static objects. This may
already provide valuable insight into the thermodynamics of 
QCD at non-zero baryon number density already from quenched QCD.

In the canonical approach simulations at non-zero $B$ can be performed 
on relatively large lattices and the use of baryon number
densities up to a few times nuclear matter density is possible
\cite{Eng99b,Kac99}.
The simulations performed so far in the static limit 
show the basic features expected at
non-zero density. As can be seen from the behaviour of the Polyakov loop
expectation value shown in Fig.~\ref{fig:density}
the transition region gets shifted to smaller temperatures
(smaller coupling $\beta$). The broadening of the transition region 
suggests a smooth crossover behaviour at non-zero density. However, 
in a canonical simulation it also may indicate
the presence of a region of coexisting phases and thus would signal
the existence of a $1^{st}$ order phase transition. This deserves
further analysis.

Even more interesting is the behaviour of the heavy quark
free energy in the low temperature phase. As shown in the right frame of
Fig.~\ref{fig:density} the free energy does get screened at
non-vanishing number density. The influence of static quark sources on
the heavy quark free energy is similar in magnitude to the 
screening (string breaking) seen in QCD simulations at finite temperature 
in the low temperature hadronic phase (see Fig.~\ref{fig:pot_pg_nf3}). 
At non-zero baryon number density we thus may expect similarly
strong medium effects as at finite temperature. 

\section{Conclusions}

We have given a brief introduction into the lattice formulation of
QCD thermodynamics and presented a few of the basic results on
the QCD equation of state, the critical parameters for the transition
to the QCD plasma phase and properties of this new phase of matter.

The thermodynamics of the heavy quark mass limit is quite well
under control; we know the equation of state and the transition
temperature with an accuracy of a few percent.  We now also have 
reached a first quantitative understanding of QCD with light quarks, 
which at present still corresponds to a world in which the pion would 
have a mass of about (300-500)~MeV. This still is too heavy to become 
sensitive 
to details of the physics of chiral symmetry breaking. Nonetheless,
lattice calculations performed with different lattice fermion
formulations start to produce a consistent picture for the
quark mass dependence of the equation of state as well as the influence 
of the number of light flavours on the phase transition and they yield
compatible results for the transition temperature. In these 
calculations we learn to control the systematic errors inherent
to lattice calculations performed with a finite lattice cut-off
and start getting control over the effects resulting from the 
explicit breaking of continuum symmetries in the fermion sector.
With these experience at hand we soon will be able to study
the thermodynamics of QCD with a realistic spectrum of light up 
and down quarks and a heavier strange quark.

\section*{Acknowledgement}

I would like to thank the organizers of the 40th Schladming Winter 
School on ``Dense Matter'' for arranging a very stimulating 
topical meeting. Furthermore, I would like to thank all my
colleagues in Bielefeld, in particular J\"urgen Engels, 
Edwin Laermann, Bengt Petersson and Helmut Satz,  for the
extremely fruitful collaboration I had with them for many
years on most of the topics discussed in these lectures.

\section*{Appendix: Improved Gauge and Fermion Actions}

\noindent
{\bf Improved gauge actions}

\vspace{0.1cm}
\noindent
When formulating a discretized version of QCD one has a great deal of
freedom in choosing a lattice action. Different formulations may differ
by
sub-leading powers of the lattice cut-off, which vanish in the continuum
limit. This has, for instance, been used by Symanzik to systematically
improve scalar field theories \cite{Sym83} and has then been applied
to lattice regularized $SU(N)$ gauge theories \cite{Wei83,Lue85}. In
addition to the elementary plaquette term appearing in the standard 
Wilson formulation of lattice QCD larger loops can be added to the action 
in such a way that the leading ${\cal O} (a^2g^0)$ deviations from the 
continuum formulation are eliminated and corrections only start in 
${\cal O} (a^4g^0, a^2g^2)$. A simple class of improved actions is, for 
instance, obtained by adding planar loops of size $(k,l)$ to the standard 
Wilson action (one-plaquette action) \cite{Bei96}. The simplest 
extension of the Wilson one-plaquette action thus is to include 
an additional contribution from a planar six-link Wilson loop,
\begin{equation}
W_{n, \mu\nu}^{(1,2)} = 1 - \frac{1}{6}\; {\rm Re}\; \biggl(
\loOp \; + \; \lOop \biggr)_{n,\mu \nu} \quad  .
\end{equation}
Combining this six-link contribution with the four-link plaquette 
term in a suitable way one can eliminate the leading ${\cal O}(a^2)$
corrections and arrives at a formulation that reproduces the
continuum action up to ${\cal O}(a^4)$ corrections at least on the
classical level at ${\cal O}(g^0$),

\begin{equation}
\beta S_G = \beta \sum_{\scriptstyle n \atop\scriptstyle
0\le\mu<\nu\le3}
c_{1,1} W_{n,\mu\nu}^{(1,1)} \; + \; c_{1,2} W_{n,\mu\nu}^{(1,2)} 
\quad ,
\label{s12}
\end{equation}
with $c_{1,1} = 5/3$ and $c_{1,2} = -1/6$. 
We call this action the tree-level improved $(1\times 2)$-action.
It may be further improved perturbatively by eliminating the leading 
lattice cut-off effects also at ${\cal O}(g^2)$, {\it i.e.} 
$c_{i,j} \Rightarrow c_{i,j}^{(0)} + g^2 c_{i,j}^{(1)}$, or 
by introducing non-perturbative modifications. A well-studied
gluon action with non-perturbative corrections is the RG-improved
action introduced by Y. Iwasaki \cite{Iwasaki}. This {\it RG-action} 
also has the structure of Eq.~\ref{s12} but with coefficients 
$c_{1,1}^{RG} = 3.648$ and $c_{1,2}^{RG} = -0.662$. Of course, this
action will still lead to ${\cal O}(a^2)$ corrections in the ideal
gas limit. The $N_\tau$-dependence of cut-off effects resulting 
from these actions is shown in Fig.~\ref{fig:improvedideal}. 
In Section 6 we also show some results from calculations with a 
tadpole improved actions \cite{Lep93}. This non-perturbative 
improvement amounts to $c_{1,1}^{\rm tad} \equiv c_{1,1}$ and a 
replacement of 
$c_{1,2}$ by $c_{1,2}^{\rm tad} = 1/6u_0^2(\beta)$
where,
\begin{equation}
u_0^4 = {1 \over 6 N_\sigma^3 N_\tau}~
\langle \sum_{x, \nu > \mu} (1-W^{1,1}_{\mu, \nu} (x))~\rangle \quad .
\label{u0}
\end{equation} 
In the ideal gas limit this action still has the same cut-off dependence
as the tree-level improved $(1\times 2)$ action.


\vspace{0.5cm}
\noindent
{\bf Improved staggered fermion actions}

\vspace{0.1cm}
\noindent
When discussing the improvement of fermion actions there are 
at least two aspects one has to take into account. On the one
hand one faces problems with cut-off effects similar to the
pure gauge sector; on the tree level the standard Wilson and 
Kogut-Susskind discretization schemes introduce ${\cal O}(a^2)$
which will influence the short distance properties of physical
observables. On the other hand also the global symmetries of the 
continuum Lagrangian are explicitly broken at non-zero
lattice spacing. This influences the long distance properties 
of these actions, e.g. the light particle sector (Goldstone modes)
of the lattice regularized theory. Both aspects are of importance
for thermodynamic calculations. The latter problem certainly
is of importance in the vicinity of the QCD phase transition
while the former will show up when analyzing the 
high temperature limit of the equation of state.

In the case of staggered fermions both problems have been
addressed and schemes have been developed that lead to
a reduction of cut-off effects at short distances, {\it i.e.}
high temperature, and also allow to reduce the explicit
flavour symmetry breaking of the staggered discretization scheme.  

A particular form of improved action used in recent calculations
of the Bielefeld group is a staggered fermion action, which
in addition to the standard one-link term includes a set of
bended three-link terms,

\begin{eqnarray}
\lefteqn{ S_F (m_{q}) ~=~ c_1^F S_{1-link,fat} (\omega) +
c_3^F S_{3-link}+  m_{q}\sum_x
\bar{\chi}_x^f \chi_x^f  ~~~~~~~~~~~~~~~~~~~~~~~}\nonumber \\
&\equiv& \sum_x \bar{\chi}_x^f~\sum_\mu ~ \eta_\mu(x) ~ \Bigg(
{3\over 8}~\Bigg[ \alink~ +~ \omega~~\sum_{\nu \ne \mu}~~
\alinkfat\Bigg] \nonumber \\[4mm]
& & + {1\over96}~\sum_{\nu\ne \mu} ~\Bigg[ \blinkbd + \blinkbc ~+
 \blinkba + \blinkbb \Bigg] \Bigg) \chi_y^f \nonumber \\[13mm]
& & + m_{q}  \sum_x~\bar{\chi}_x^f \chi_x^f  \quad .
\label{sf}
\end{eqnarray}
Here $\eta_\mu(x) \equiv (-1)^{x_0+..+x_{\mu-1}}$ denotes the
staggered fermion phase factors. Furthermore, we have made explicit
the dependence of the fermion action on different
quark flavours $q$, and the corresponding bare quark masses $m_q$,
and give an intuitive graphical representation of the action.
The tree level coefficients $c_1^F$ and $c_3^F$ appearing in $S_F$
have been fixed by demanding rotational invariance
of the free quark propagator at ${\cal O}(p^4)$  (``p4-action'')
\cite{Hel99}.
In addition the 1-link term of the fermion action has been modified by
introducing ``fat'' links \cite{Blu97} with a weight $\omega = 0.2$.
The use of fat links does lead to a reduction of the flavour symmetry
breaking close to $T_c$ and at the same time does not modify the
good features of the p4-action at high temperature, {\it i.e.} it 
does not modify the cut-off effects at tree level and has
little influence on the cut-off dependence of bulk thermodynamic
observables at ${\cal O}(g^2)$ in the high temperature phase 
\cite{Hel99}.
Further details on the definition of the action are given
in \cite{Hel99}.

We refer to this action with a fat 1-link term combined with the tree 
level improved gauge action as {\it the p4-action}. 
The $N_\tau$-dependence of cut-off effects resulting 
from this action is shown in Fig.~\ref{fig:improvedideal}.

%
 
\end{document}